\documentclass[11pt,a4paper]{article}
\pdfoutput=1
\usepackage{jheppub}
\usepackage[latin1]{inputenc}
\usepackage[T1]{fontenc}
\usepackage[american]{babel}
\usepackage{amsmath}
\usepackage{amsthm}
\usepackage{amsfonts}
\usepackage{amssymb}
\usepackage{array}
\usepackage{color}
\usepackage{dsfont}
\usepackage{lscape}
\usepackage{slashed}
\usepackage{hyperref}
\usepackage{mciteplus}

\DeclareMathOperator{\Trace}{\operatorname{Tr}}
\DeclareMathOperator{\trace}{\operatorname{tr}}

\DeclareMathOperator{\D}{\mathcal{D}}

\newcommand{\eq}{eq.~}
\newcommand{\eqs}{eqs.~}
\newcommand{\fig}{fig.~}

\newcommand{\app}{app.~}

\newcommand{\secref}[1]{sec.\ \ref{#1}}
 
\renewcommand{\Re}{\operatorname{Re}}

\renewcommand{\d}{\textmd{d}}
\newcommand{\be}{\begin{equation}}
\newcommand{\ee}{\end{equation}}
\newcommand{\Z}{\mathcal{Z}}
\newcommand{\expv}[1]{\left \langle #1 \right \rangle}

\newcommand{\ve}[1]{{\mathbf #1}}
\def\pe{_\perp}
\def\pa{_\parallel}

\def\seff{S_{{\rm eff}}}

\def\ba{B}
\def\ea{E}
\def\aa{A}
\def\fa{F}

\def\bna{\mathcal{B}}
\def\ena{\mathcal{E}}
\def\ana{\mathcal{A}}
\def\fna{\mathcal{F}}
\renewcommand{\Z}{\mathcal{Z}}
\renewcommand{\D}{\mathcal{D}}

\def\dena{A(\mathcal{E})}
\def\dbna{A(\mathcal{B})}

\def\f{\mathcal{C}}
\def\F{\mathcal{C}}

\def\X{-\Delta I_g^{\rm imp}}

\title{Magnetic field-induced gluonic (inverse) catalysis and pressure (an)isotropy in QCD}

\author[1,2]{G.~S.~Bali,}
\author[1]{F.~Bruckmann,}
\author[1]{G.~Endr\H{o}di,}
\author[1]{F.~Gruber,}
\author[1]{A.~Sch\"afer}

\affiliation[1]{Institute for Theoretical Physics, Universit\"at Regensburg, D-93040 Regensburg, Germany.}
\affiliation[2]{Tata Institute of Fundamental Research, Homi Bhabha Road, Mumbai 400005, India.}

\emailAdd{gunnar.bali@physik.uni-regensburg.de}
\emailAdd{falk.bruckmann@physik.uni-regensburg.de}
\emailAdd{gergely.endrodi@physik.uni-regensburg.de}
\emailAdd{florian.gruber@physik.uni-regensburg.de}
\emailAdd{andreas.schaefer@physik.uni-regensburg.de}

\abstract{We study the influence of strong external magnetic fields on gluonic and fermionic observables in the QCD vacuum at zero and nonzero temperatures,
via lattice simulations with $N_f=1+1+1$ staggered quarks of physical masses.
The gluonic action density is found to undergo magnetic catalysis at low temperatures and inverse magnetic catalysis near and above the
transition temperature, similar to the quark condensate.
Moreover, the gluonic action develops an anisotropy: the
chromo-magnetic field parallel to the external field is enhanced,
while the chromo-electric field in this direction is suppressed.
We demonstrate that the same hierarchy is obtained using the Euler-Heisenberg
effective action. 
Conversely, the topological charge
density correlator does not reveal a
significant anisotropy up to magnetic fields $eB\approx 1\textmd{ GeV}^2$.
Furthermore, we show that the pressure remains isotropic even for nonzero magnetic fields, if it is defined through
a compression of the system at fixed external field.
In contrast, if the flux of the field is kept fixed during the compression -- which is the situation realized in the lattice simulation -- the pressure develops an anisotropy.
We estimate the quark and gluonic contributions to this anisotropy, and relate them to the magnetization of the QCD vacuum. After performing electric charge renormalization, we obtain an estimate for the magnetization, which indicates that QCD is paramagnetic.}


\begin{document}
\maketitle

\section{Introduction}

Strong magnetic fields are an interesting probe of Quantum Chromodynamics (QCD). They are relevant in at least three strongly interacting physical systems: the early universe, magnetars and non-central heavy ion collisions~\cite{Vachaspati:1991nm,Duncan:1992hi,Skokov:2009qp}. Moreover, external magnetic fields challenge our theoretical understanding of non-perturbative phenomena in QCD. The chiral magnetic effect~\cite{Kharzeev:2007jp,Fukushima:2008xe}, where the magnetic field may couple to topological features of QCD (on an event-by-event basis), has triggered a lot of activity in this direction.

A rather robust effect is the enhancement of the quark condensates by the magnetic field $B$, and, thus, the generation of dynamical mass, the so-called magnetic catalysis \cite{Gusynin:1995nb}. The perturbative approach to this 
mechanism (and also to other effects)
focuses on the Landau levels, i.e., on the eigenvalues of the Dirac equation 
in constant external magnetic backgrounds. The lowest Landau level has vanishing eigenvalue and a degeneracy proportional to the magnetic flux~\cite{Landau:1930}. Therefore, the magnetic field increases the density of small eigenvalues and thus, via the Banks-Casher relation~\cite{Banks:1979yr}, the quark condensate. The enhancement of the condensate at zero temperature is also directly related to the positivity of the scalar QED $\beta$-function~\cite{Endrodi:2013cs}.
Magnetic catalysis has been confirmed in various non-perturbative approaches to QCD, as well as in ab-initio lattice simulations~\cite{D'Elia:2010nq,Bruckmann:2011zx,Bali:2011qj,Bali:2012zg,Ilgenfritz:2012fw}. For a recent review see, e.g., ref.~\cite{Shovkovy:2012zn}.
At finite temperature, however, \emph{inverse} magnetic catalysis was found for physical quark masses~\cite{Bali:2011qj,Bali:2012zg}, manifested by a reduction of the condensate with growing $B$. This effect reduces the transition temperature $T_c$ of the QCD crossover~\cite{Bali:2011qj}.
Inverse magnetic catalysis may be attributed to the sea quark back-reaction being particularly effective near the transition~\cite{Bruckmann:2013oba}.

The presence of a uniform homogeneous magnetic field breaks the isotropy of space. The Landau level eigenfunctions are plane waves in the direction of the magnetic field, whereas they are (localized) harmonic oscillator eigenfunctions perpendicular to $B$. Accordingly, in the interacting case, the Dirac eigenmodes are expected to be elongated~\cite{Basar:2011by} along the magnetic field, and for asymptotically large $B$, the dynamics of the system effectively reduces
to $1+1$ dimensions. The anisotropy will result in nonzero expectation values of certain observables that at zero field vanish due to rotational symmetry. A prominent example is the magnetic susceptibility of the QCD vacuum~\cite{Ioffe:1983ju}, the proportionality factor between the spin polarization $\langle\bar{\psi}\sigma_{xy}\psi\rangle$, the external magnetic field (directed along the positive $z$ axis) $B\equiv B_z=F_{xy}$ and the condensate $\langle\bar{\psi}\psi\rangle$. Its value has recently been determined using lattice simulations at physical light and strange quark masses, in the continuum limit~\cite{Bali:2012jv} (for quenched results see~\cite{Buividovich:2009ih,Braguta:2010ej}).

In this paper, we mainly focus on gluonic quantities. 
Since only the quarks are electrically charged, the effect of magnetic
fields on the gluons may at first seem secondary. Nevertheless,
we find the \emph{total gluonic action} and the \emph{anisotropic
magnitudes of the field strength components}
to reveal a pronounced dependence on the magnetic field. 
Interestingly, the gluonic action behaves quite similar to the condensate. At zero temperature it is increased by the magnetic field, around the transition it is non-monotonic, whereas above $T_c$ it undergoes inverse magnetic catalysis. Note that the gluonic action is in some sense tied to the quark condensate: the sum of both enters the interaction measure (trace anomaly). We will exploit this relation
to improve the definition of the gluonic action at our nonzero
lattice spacings.

Our second objective is to determine the anisotropy between the (squared) gluonic field strengths parallel and perpendicular to the external field. 
This anisotropy was first measured on the lattice by the Berlin group~\cite{Ilgenfritz:2012fw}, for $\mathrm{SU}(2)$. The following hierarchy was found: the chromo-{\it magnetic} field parallel to $B$ has a larger, and the chromo-{\it electric} field parallel to $B$ a smaller expectation value than the components perpendicular to the magnetic field. The latter all share the same expectation value at zero temperature due to the remaining symmetry (at nonzero temperatures, an additional anisotropy is induced between the chromo-magnetic and chromo-electric fields). We confirm this hierarchy for real QCD with three quark flavors of physical masses.
For a comparison, we give the Euler-Heisenberg effective action for small color fields in a constant external magnetic field background. It reveals the same hierarchy. 

We also determine the anisotropies of the fermionic part of the QCD action, and discuss how these -- together with the gluonic anisotropies -- are related to the diagonal components of the energy-momentum tensor. The subtle question, whether the hydrodynamic pressure in the presence of magnetic fields is isotropic or anisotropic is the topic of an ongoing discussion, see e.g. refs.~\cite{Blandford:1982,Ferrer:2010wz,Potekhin:2011eb,
Ferrer:2011ze,Strickland:2012vu}, and is discussed in \secref{sec isoaniso}. This question is highly relevant, for instance, for density fluctuations in the early universe~\cite{Kandus:2010nw}, for magnetized neutron stars~\cite{Duncan:1992hi} and for the elliptic flow observed in heavy-ion collisions~\cite{Muller:2012zq,*Schenke:2012wb,*Heinz:2013th}. 
We show that the pressure is isotropic, if defined via the compression of the system with the magnetic field kept fixed, whereas it is anisotropic if the flux of the field is kept constant during the compression.
We demonstrate how the latter anisotropy can be used to obtain the magnetization of the QCD vacuum, and give the first estimate on the renormalized magnetization from lattice simulations.

For the third gluonic quantity under study, the two-point correlator of the topological charge density, 
we do not find a significant anisotropy for the applied range of magnetic fields. One might have expected an anisotropy mediated by elongated quasi-zero modes, given the strong link between topology and chirality. Therefore, also this null result contains interesting information about the response of the QCD vacuum to a magnetic field.

This paper is organized as follows. In the next section
we define our observables in the continuum and on the lattice, and discuss their renormalization. In \secref{sec results} we briefly describe our simulation setup and present our main numerical results. Subsequently,
in \secref{sec isoaniso},
we address the anisotropy in the pressure and relate this to the QCD magnetization. This is
followed by our conclusions in \secref{sec conclusions}.
The article is augmented by four appendices that contain more
detailed derivations and the Euler-Heisenberg effective action for our setting.

\section{Observables}
\label{sec observables}

On the lattice, observables are given in terms of the partition function, which is written as the functional integral over the gauge links $U$,
\be
\Z = \int \D U e^{-\beta S_g} \, \prod_f \det\left[a\slashed{D}(\Phi) + am_f\right],
\label{eq:partfunc}
\ee
where $f=u,d,s$ labels the quark flavors, $a$ is the lattice spacing, and the parameters are the inverse gauge coupling $\beta\equiv 6/g^2$, the lattice quark masses $am_f$ and the flux $\Phi$ of the external magnetic field $eB$, which couples to the quark determinants through the electric charges $q_f/e$ of the quarks, see below. Here $e>0$ is the elementary charge. The quark condensates and the gluonic action are given by the partial derivatives with respect to the lattice parameters,
\be
\expv{\bar{\Psi}_f\Psi_f} = \frac{\partial\log \Z}{\partial (am_f)},\quad\quad
\expv{\bar\psi_f\psi_f} = \frac{a^{-3}}{N_s^3N_t} \expv{\bar\Psi_f\Psi_f}, \quad\quad
\expv{S_g} =-\frac{\partial\log \Z}{\partial \beta}, \quad\quad
\expv{s_g} = \frac{a^{-4}}{N_s^3N_t} \expv{S_g},
 \label{eq_the_quantities}
\ee
where we also defined the corresponding densities in physical units. 
Here, $N_s$ and $N_t$ are the number of lattice sites in the spatial and temporal directions, respectively, and $a$ is the lattice spacing. In terms of these, the three-volume and the temperature read $V=(N_sa)^3$ and $T=(N_ta)^{-1}$.
For convenience, we define for any observable $\mathcal{O}$,
\be
\label{eq:delta}
 \Delta \mathcal{O}= \left.\expv{\mathcal{O}}\right|_{e\ba} - \left.\expv{\mathcal{O}}\right|_{0},
\ee
as the difference of expectation values
with and without the external magnetic field.

\subsection{Interaction measure}

The gauge action density and the quark condensate densities enter the interaction measure (trace anomaly),
\be
I = -\frac{T}{V}\frac{\d \log \Z}{\d \log a},
\label{eq:tradef}
\ee
an observable relevant for the QCD equation of state. We consider the gluonic contribution to $\Delta I$, which -- after the improvement procedure discussed in \app\ref{sec imp} -- reads
\begin{align}
 \X=&\,-\frac{\partial \beta}{\partial \log a}\Delta s_g+\sum_f\left(\frac{\partial \log(am_f)}{\partial \log a}-1\right)m_f \Delta\bar{\psi}_f\psi_f.
 \label{eq_the_long_name_quantity}
\end{align}
The first term on the right hand side of \eq(\ref{eq_the_long_name_quantity}) is the direct gluonic contribution, see, e.g., ref.~\cite{Borsanyi:2010cj}. 
Since the additive divergences are canceled in $\Delta s_g$ (see the discussion in apps.~\ref{sec renorm} and \ref{sec imp}), this first term already has a well-defined continuum limit (at $B=0$, typically differences in the temperature are utilized to achieve this).
The leading lattice discretization error, however, is of 
$\mathcal{O}(1/\log a)$ and, therefore, vanishes very slowly.
The second, fermionic term in \eq(\ref{eq_the_long_name_quantity}) does not contribute to the continuum limit, but it cancels the logarithmic discretization error, thereby improving the convergence to $\mathcal{O}(a^2)$, see \app\ref{sec imp}. The improvement is evident from our numerical data, see \fig\ref{fig impr}. The procedure is equivalent to multiplying the result by a finite renormalization constant $1+\mathcal{O}(1/\beta)$ that we determine non-perturbatively, using the line of constant physics (LCP) for the action that we employ~\cite{Borsanyi:2010cj}. 
To our best knowledge, such an $\mathcal{O}(1/\beta)$-improvement of a
gluonic quantity by a fermionic quantity has not been considered in the literature previously. 

\subsection{Anisotropies}
\label{sec observables gluonic aniso}

The continuum counterpart of the gluonic action in Euclidean space is written as
\be
\beta S_g \;\leftrightarrow\; \frac{1}{2g^2} \int \d^4 x \trace \fna_{\mu\nu}^2(x)
 =\frac{1}{g^2} \int \d^4 x\trace\left[{\boldsymbol{\ena}}^{2}(x)+{\boldsymbol{\bna}}^{2}(x)\right].
\label{eq_gluonic_action_cont}
\ee
The field strength is defined in terms of the $\mathrm{SU}(3)$ gauge potential $\ana_\mu$ as $\fna_{\mu\nu}=\partial_\mu\ana_\nu-\partial_\nu\ana_\mu+i[\ana_\mu,\ana_\nu]$, and consists of chromo-electric $\ena_i=\fna_{4i}$ and chromo-magnetic components $\bna_i=\epsilon_{ijk}\fna_{jk}/2$.
We use calligraphic letters to denote the non-Abelian $\mathrm{SU}(3)$ fields, to distinguish these from the external Abelian field\footnote{Note that the Euclidean $\ena^2$ ($E^2$) turns into $-\ena^2$ ($-E^2$) in Minkowski space-time, whereas the sign of the squared magnetic fields remain the same.}. The full covariant derivative reads $D_{\mu,f}=\partial_\mu+i\ana_\mu+i q_f \aa_\mu$. 
Without loss of generality, we will take the external magnetic field $\ba$ to point in the $z$-direction. The simplest gauge field to realize this (in infinite volume) is $\aa_y=\ba x$, $\aa_\mu=0$ ($\mu\neq y$).

The translation of these quantities to the lattice discretization is straightforward. For the gauge action we use the tree-level improved Symanzik
action \cite{Weisz:1982zw},
\be
S_g  = S_g^{\rm Sym} = \sum_{\mu<\nu} \sum_n \frac{1}{3}\Re \trace P_{\mu\nu}(n),
\label{eq:sgsym}
\ee
where $P_{\mu\nu}(n)$ denotes a sum of gluonic loops
lying in the $\mu$-$\nu$ plane, see \eq(\ref{eq:weisz}), and $n$ runs over lattice sites. Therefore, $S_g$ is readily decomposed into planar components and, therefore, into squared traces of the chromo-electric and chromo-magnetic field strengths, according to \eq(\ref{eq_gluonic_action_cont}),
\be
\trace\ena_i^2(n) =  2 \Re \trace P_{4i}(n),
\quad\quad\quad
\trace\bna_i^2(n) =  2\sum_{j<k}|\epsilon_{ijk}| \Re \trace P_{jk}(n).
\label{eq:compgact}
\ee
In the following, the components in the direction of the external field $B\parallel z$
are denoted as parallel, whereas the $x$ and $y$ components as perpendicular,
\begin{equation}
 \ena\pa^2=\ena_z^2,\quad\quad\quad
 \bna\pa^2=\bna_z^2,\quad\quad\quad
 \ena\pe^2=\frac{\ena_x^2+\ena_y^2}{2},\quad\quad\quad
 \bna\pe^2=\frac{\bna_x^2+\bna_y^2}{2}.
\label{eq:ebpp}
\end{equation}
We define the anisotropies as the densities of the expectation values of differences between these components,
\begin{equation}
 \dena = \frac{T}{V}\expv{ \frac{\beta}{6}\sum_n\left( \trace\ena\pe^2(n) - \trace\ena\pa^2(n) \right)}, \quad\quad\quad
 \dbna = \frac{T}{V}\expv{ \frac{\beta}{6}\sum_n \left( \trace\bna\pe^2(n) - \trace\bna\pa^2(n) \right)},
\label{eq def delta EB}
\end{equation}
which are intensive, gauge invariant quantities. The factor of $\beta/6=g^{-2}$ is included here to define the anisotropy of the total action, \eq(\ref{eq_gluonic_action_cont}). At $B=0$, the spatial isotropy of the system is restored, and both $\dena$ and $\dbna$ vanish.

Now we turn to the anisotropy induced in the fermionic part of the action, $\bar\Psi_f \slashed{D}\Psi_f$. Even though the Dirac structure of $\slashed{D}$ in the staggered discretization is related non-trivially to the continuum structure, the fermionic action can still be unambiguously separated into four terms, corresponding to space-time directions $\mu$. Each quark flavor $f$ gives a contribution,
\be
\f_{\mu,f} = \bar\Psi_f \slashed{D}_\mu \Psi_f,
\ee
where the $\slashed{D}_{\mu}$ are defined in \eq(\ref{eq:diracop})
for staggered quark actions.
In analogy to \eqs(\ref{eq:ebpp}) and (\ref{eq def delta EB}), we define the parallel and perpendicular components and the anisotropy as
\begin{equation}
 \F_{\parallel,f}=\f_{z,f},\quad\quad\quad 
 \F_{\perp,f}=\frac{\f_{x,f}+\f_{y,f}}{2},\quad\quad\quad
 A(\F_f)=\frac{T}{V}\expv{\F_{\perp,f}-\F_{\parallel,f}}.
\label{eq def delta ferm}
\end{equation}
Again, $A(\F_f)$ vanishes in the absence of the external field.
In \secref{sec isoaniso}, we will discuss the renormalization properties of these anisotropies and their relationship to the QCD magnetization.

\subsection{Topological charge density and its correlator}

The density of topological charge in the continuum reads
\begin{equation}
 q(x)=\frac{1}{32\pi^2}\, \epsilon_{\mu\nu\rho\sigma}\trace \fna_{\mu\nu}(x)\fna_{\rho\sigma}(x)
 =\frac{1}{8\pi^2}\,\trace\boldsymbol{\ena}(x)\boldsymbol{\bna}(x)\,.
\end{equation}
This definition is not suited for a direct application on the lattice, as the latter is discrete, and short range quantum fluctuations would dominate the signal. Therefore, we 
apply 5 steps of improved stout smearing~\cite{Moran:2008ra} (filtering) with the standard smearing parameters $\rho=0.06$ and $\epsilon=-0.25$. The 3-loop-improved field strength tensor~\cite{BilsonThompson:2002jk} is then measured and used in the above formula to define $q(x)$. This technique has been applied in various studies of topological properties~\cite{Ilgenfritz:2008ia,Bruckmann:2011ve}, since it gives a topological density similar to that from the fermionic definition containing all eigenmodes of the overlap operator~\cite{Niedermayer:1998bi,Horvath:2002yn}. Less filtering leads to ambiguous results for the improved field strength, whereas stronger filtering
eventually washes out the topological structures.   

The topological density is a CP-odd variable and, thus, vanishes on average in the QCD vacuum. This remains true in the presence of an external magnetic field (only in combination with an additional external electric field $\mathbf{\ea}$ a second CP-odd variable exists, $\mathbf{\ea}\mathbf{\ba}$, that could couple to $q$, see, e.g., refs.~\cite{Rafelski:1998tc,Elze:1998wm,D'Elia:2012zw}).

We will consider the two-point correlator of the topological density,
$ \langle q(x)q(x')\rangle$. It is a CP-even quantity that, at $B=T=0$, depends only on the absolute value of the distance of the two points, due to translational and rotational invariance. At $B>0$, rotational symmetry is lost, and therefore the correlator may be different if the four-dimensional difference vector $x-x'$ is perpendicular to the magnetic field or parallel to it. At $T>0$, the only parallel direction is the $z$-direction, whereas at $T=0$, the parallel directions also include the $t$-direction, 
\begin{equation}
 \left.\begin{array}{l}
  \langle q(0)q(r)\rangle_\perp \\ 
  \langle q(0)q(r)\rangle_\parallel \\ 
  \langle q(0)q(r)\rangle_{{\rm full}}
 \end{array}\right\}\equiv\langle q(x)q(x')\rangle\quad \mbox{ with } 
 |x-x'|=r\mbox{ and }
 x-x' \in \left\{
 \begin{array}{l}
  (x,y) \mbox{-plane}\\ (z,(t)) \mbox{-plane}\\ \mbox{arbitrary}
 \end{array}\right.\,.
\label{eq_top_corr_def}
\end{equation}
We consider the difference of these correlators to quantify the anisotropy in the topological sector.

The integral of the full correlator over $r$ gives the topological susceptibility $T/V \cdot\langle Q^2\rangle$, and is thus positive. At intermediate distances, however, 
the correlator becomes negative, with the position of the zero proportional to the lattice spacing. This is clear from the continuum limit, in which the correlator is negative for all distances apart from zero, where it contains a contact term~\cite{Seiler:1987ig}. The scaling of the position of the zero has turned out to be very similar for many current fermion discretizations~\cite{Bruckmann:2011ve}. 
We will focus on intermediate distances $r/a=2 \ldots 4$. Note that, due to the restriction of $r$ to two-dimensional hyperplanes, less discrete distances are available for the perpendicular and parallel correlators than for the full propagator, resulting in lower effective statistics.

Finally, we remark that the use of smearing and of the improved field strength definition may affect the anisotropy of the correlator, as both techniques effectively amount to an averaging over space-time regions in a spherically symmetric way.
For the mild averaging that we employ, however, these regions do not overlap strongly and,
thus, the correlators defined above still contain the information about anisotropies
in the topological charge. 

\section{Results I: gluonic and fermionic observables}
\label{sec results}

Our measurements have been performed on the same configurations as used in our previous studies of magnetic fields in QCD. The configurations at zero and nonzero temperature have been generated with the tree-level improved Symanzik gauge action and $N_f=1+1+1$ stout smeared staggered fermions, at physical quark masses, for details see~\cite{Bali:2011qj,Bali:2012zg,Bali:2012jv}. The light quark masses are set equal, $m_u=m_d\equiv m_{ud}$, whereas the strange quark mass is $m_s=28.15\cdot m_{ud}$. The quark masses are tuned as a function of $\beta$ along the line of constant physics (LCP)~\cite{Borsanyi:2010cj}, which ensures that for all lattice spacings, the hadron masses are at their physical values.
The quark charges are $-q_u/2=q_d=q_s=-e/3$.

\subsection{Interaction measure}
\label{sec inter}

\begin{figure}[ht!]
\includegraphics[width=0.49\textwidth]{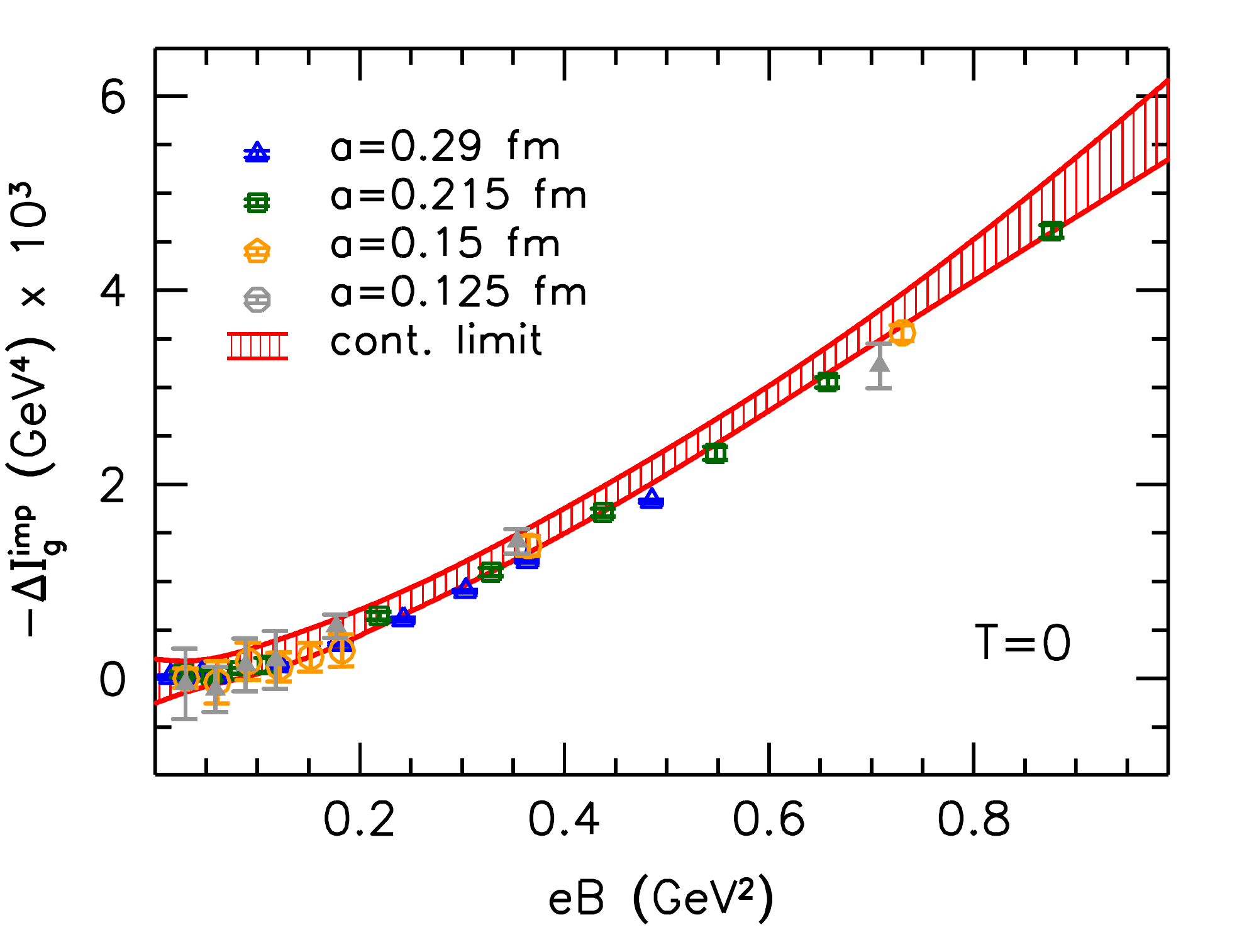}
\includegraphics[width=0.49\linewidth]{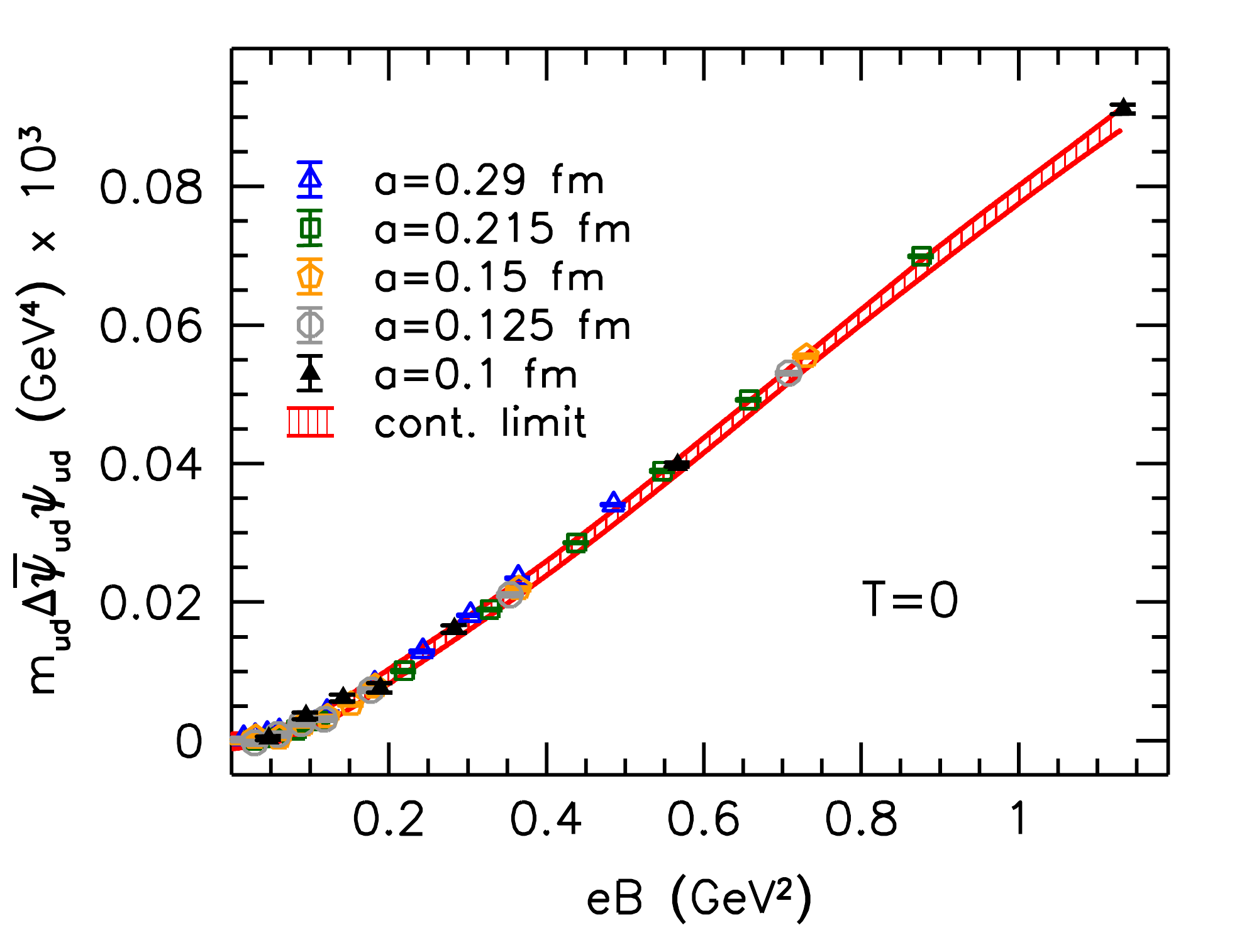}
\caption{The change in the gluonic contribution to the interaction measure, \eq(\protect\ref{eq_the_long_name_quantity}), at zero temperature (left, including the continuum limit from four lattice spacings) and
the corresponding light quark contribution $m_{ud}\Delta\bar{\psi}_{ud}\psi_{ud}$ (five lattice spacings and the continuum limit), in the same units.}
\label{fig gluonic}
\end{figure}

\noindent
We start the analysis by considering the change of the renormalized gluonic action, i.e., of the gluonic contribution to the interaction measure $\X$ of \eq(\ref{eq_the_long_name_quantity}) at zero temperature. We use four different lattice spacings with magnetic fields $eB$ up to about $1\textmd{ GeV}^2$, and perform a combined spline interpolation and continuum extrapolation to obtain the $a\to0$ limit. The results, together with the continuum limit, are shown in the left panel of \fig\ref{fig gluonic}. Clearly, $\X$ is enhanced by the magnetic field, which we interpret as {\it magnetic catalysis} for gluons. For large magnetic fields, the initial quadratic dependence seems to turn linear, just like for the quark condensate in the same system. Note that the fermionic counterpart of $\X$ is $-\Delta I_f^{\rm imp}=\sum_f m_f \Delta \bar\psi_f\psi_f$, see the discussion in \app\ref{sec imp}.
Therefore, for comparison, in the right panel of \fig\ref{fig gluonic}, we also display the change in the average light condensates, $m_{ud}\Delta\bar\psi_{ud}\psi_{ud}=m_{ud}\left(\Delta \bar\psi_u\psi_u+\Delta\bar\psi_d\psi_d\right)/2$, taken from ref.~\cite{Bali:2012zg}, and plotted in the same units as the gluonic observable in the left panel. Note that the relative errors are larger in the gluonic quantity, since here $\mathcal{O}(a^{-4})$ divergences cancel in the difference $\Delta$, whereas the leading divergence for the quark condensates is only of $\mathcal{O}(a^{-2})$~\cite{Borsanyi:2010cj}.

In absolute terms, the effect of the gluonic magnetic catalysis is by two orders of magnitude larger than that
of the light quarks. Notice however, that the natural scale of the quark condensate at $T=B=0$ is given through the Gell-Mann-Oakes-Renner relation by $M_\pi^2F^2/2\approx 6\cdot10^{-5} \textmd{ GeV}^4$ with $M_\pi$ being the pion mass and $F$ the pion decay constant, whereas conventional estimates for the gluon condensate are of the order of $10^{-2} \textmd{ GeV}^4$~\cite{Ioffe:2002be}. Therefore, the relative effect of the catalysis is stronger in the quark sector, as expected, since quarks experience the primary effect of the external magnetic field. We remark moreover, that the strange contribution is even larger, as it comes with the factor $m_s=28.15\cdot m_{ud}$.

\begin{figure}[ht!]
\includegraphics[width=0.49\linewidth]{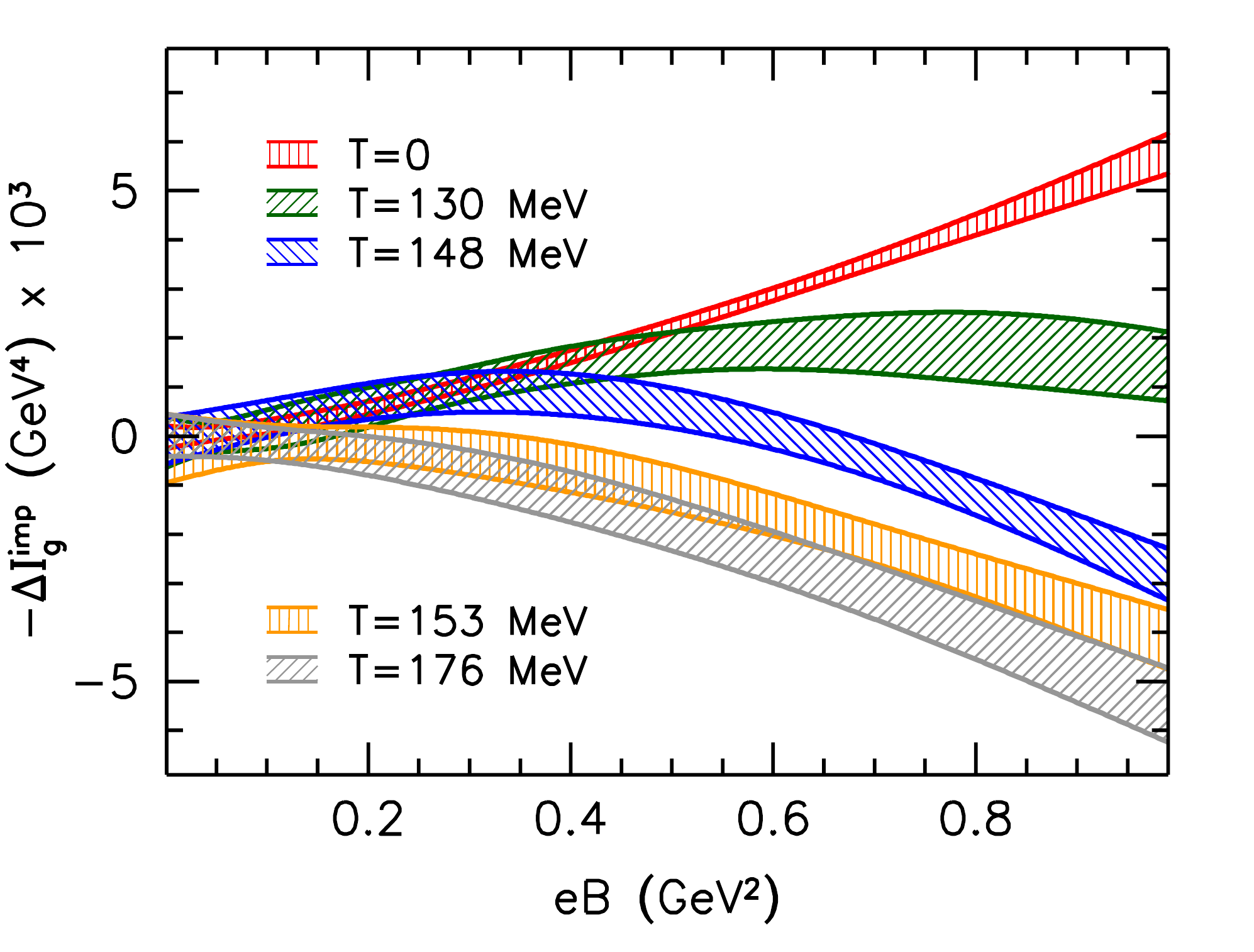}
\includegraphics[width=0.49\textwidth]{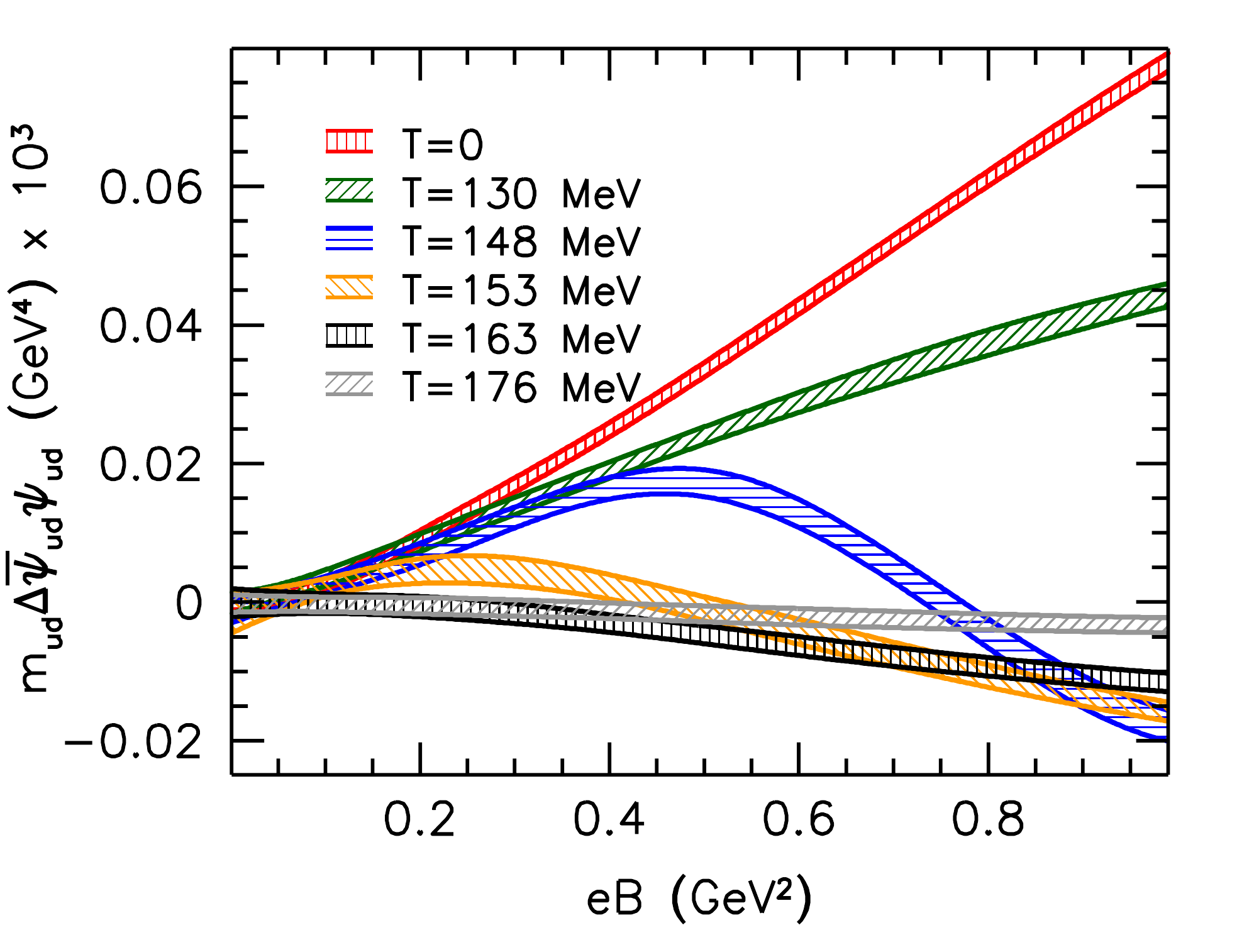}
\caption{The change in the gluonic contribution to the interaction measure at temperatures around the transition (left, continuum limits). The magnetic catalysis turns into inverse magnetic catalysis, very similar to what has been found for the quark condensate~\protect\cite{Bali:2012zg} (right panel). 
}
\label{fig gluonic2}
\end{figure}

Next, we turn to the transition region and perform a similar combined continuum extrapolation of $\X$ as we did at $T=0$, this time using three lattice spacings with $N_t=6,8$ and $10$. 
We find the behavior of $\X$ to become quite different, when the temperature approaches $T_c\approx 150 \textmd{ MeV}$, see the left panel of \fig\ref{fig gluonic2}. 
Slightly below this temperature, the dependence of the gluonic contribution to the interaction measure on the magnetic field starts to become non-monotonic. At even higher temperatures, $\X$ decreases monotonically, with magnitudes comparable to those at zero temperature: the gluons undergo {\it inverse magnetic catalysis}. Again, this behavior is very similar to that of the light quark condensate, as found in ref.~\cite{Bali:2012zg} (see the right panel of the figure), with the notable difference that gluonic inverse catalysis persists for high temperatures, whereas for $T>T_c$, the quark condensates eventually approach zero, due to chiral symmetry restoration.

\subsection{Anisotropies}
\label{sec:anisotropies}

Here, we study the individual components of the gauge action, as given by \eq(\ref{eq:ebpp}). 
We remark that for the anisotropies -- unlike in \secref{sec inter} above -- we do not perform the continuum limit, but only show the scaling tendency of the results with the lattice spacing. In order to carry out a proper continuum extrapolation, one has to subtract terms $\sim (eB)^2\log a$ that arise from charge renormalization (see \app\ref{sec renorm}). We will revisit this issue in \secref{sec isoaniso}.

\begin{figure}[ht!]
\centering
\includegraphics[width=0.49\linewidth]{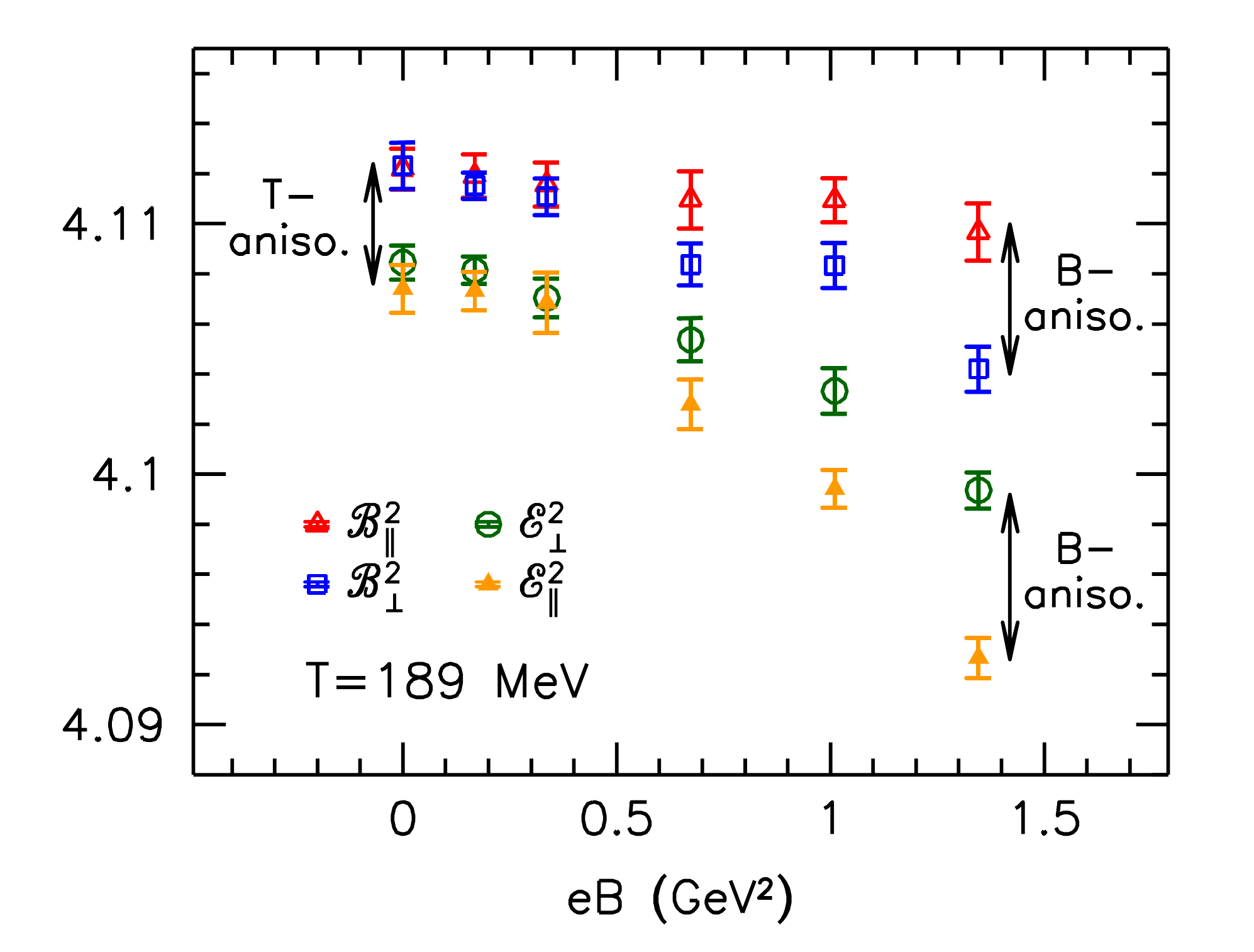}
\caption{The components $T/V\expv{\sum_n \trace \bna_i^2(n)}$ and $T/V\expv{\sum_n \trace \ena_i^2(n)}$ in lattice units $a^{-4}$, as measured on a $24^3\times 6$ lattice at a temperature $T=189 \textmd{ MeV}$. The anisotropies induced by the temperature and by the magnetic field are indicated by the arrows.}
\label{fig gluonica}
\end{figure}

First, we demonstrate the hierarchy of the gluonic components at $T>T_c$ and $B>0$, where effects from both the temperature and the magnetic field are present. 
In \fig\ref{fig gluonica}, we plot the expectation values of the
densities of the individual components
\eq(\ref{eq:ebpp}), 
as determined on our $N_t=6$ lattices. In the absence of the magnetic field, the anisotropy is induced solely by the temperature, separating the chromo-magnetic and chromo-electric components. For $B>0$, in addition the parallel and perpendicular components split, due to the spatial anisotropy induced by the magnetic field\footnote{Note that the data in \fig\ref{fig gluonica} contains an additive divergence $\sim a^{-4}$. The anisotropies induced by $T$ and $B$ (indicated by the arrows in the figure) are, however, ultraviolet finite.}. The generated hierarchy is $\big\langle{\trace\bna\pa^2}\big\rangle>\big\langle{\trace\bna\pe^2}\big\rangle>\big\langle{\trace\ena\pe^2}\big\rangle>\big\langle{\trace\ena\pa^2}\big\rangle$, similar to what was observed in the $\mathrm{SU}(2)$ theory in ref.~\cite{Ilgenfritz:2012fw}.

\begin{figure}[ht!]
\includegraphics[width=0.49\linewidth]{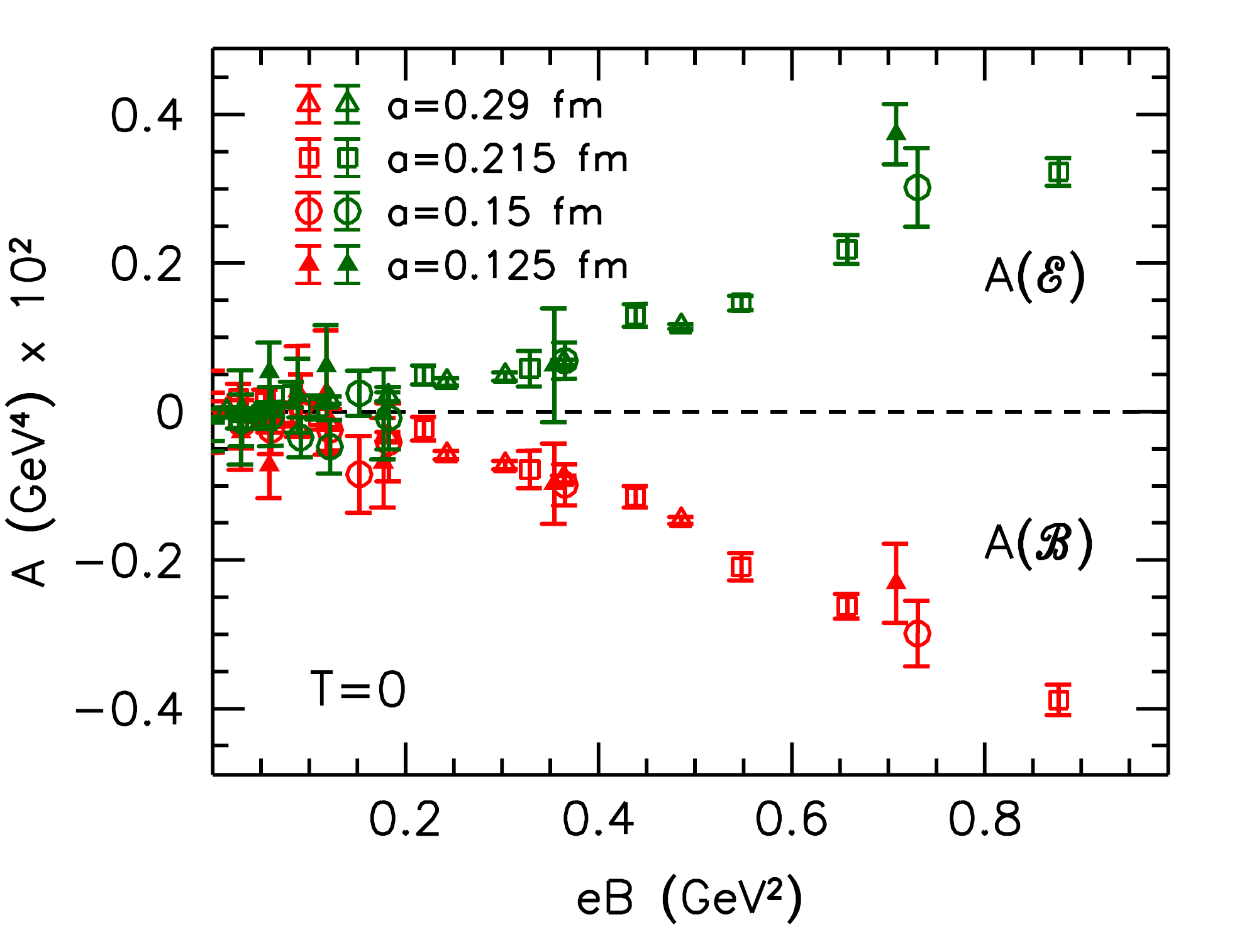}
\includegraphics[width=0.49\linewidth]{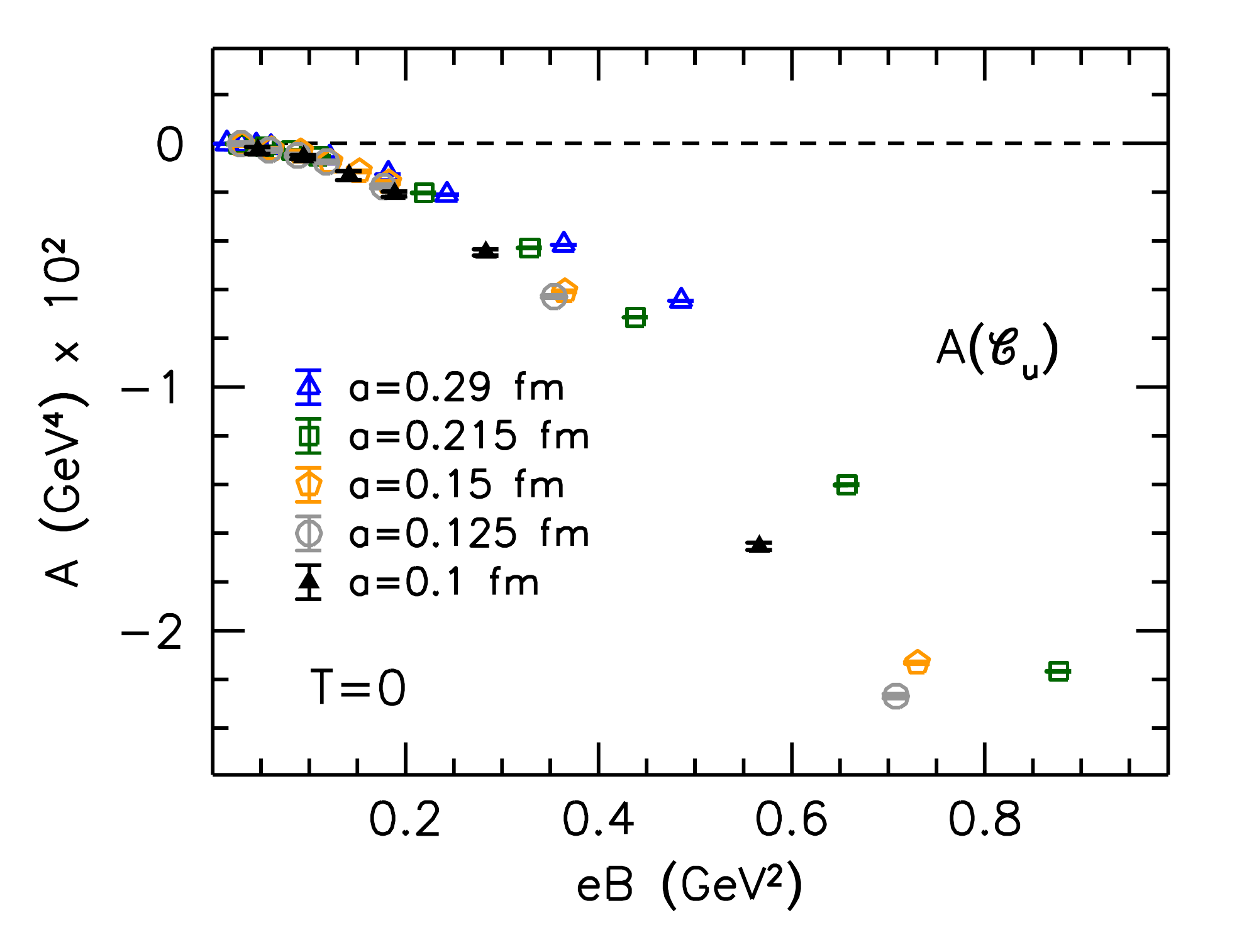}
\caption{Anisotropies in the squared field strengths, \eq(\protect\ref{eq def delta EB}), (left panel) and in the fermionic action \eq(\protect\ref{eq def delta ferm}) (right panel) at zero temperature.}
\label{fig anisotropies one}
\end{figure}

To determine the dependence of the anisotropies on the external magnetic field, in the left panel of \fig\ref{fig anisotropies one}, we plot $\dena$ and $\dbna$ (see their definition in \eq(\ref{eq def delta EB})) as functions of $eB$ at $T=0$. 
The parallel chromo-electric field is suppressed with respect to the perpendicular fields, resulting in a positive $\dena$, whereas the chromo-magnetic sector shows the opposite effect, giving a negative $\dbna$. 
This non-perturbative finding is in-line with a perturbative treatment
of the anisotropy, see the generalized Euler-Heisenberg calculation
in \app\ref{sec EH}, in particular \eq(\ref{eq fourth final}).
According to this calculation, $\trace\ena\pa^2$ increases the effective action (to bi-quadratic order in $\ba$ and $\fna$), 
and is thus suppressed. In contrast, $\trace\bna\pa^2$ reduces the action,
and is favored. This implies $\dena>0$ and $\dbna<0$, as we have found.
Furthermore, within the present statistical accuracy,
the two anisotropies have the same magnitude. 

The gluonic anisotropies do not show any significant finite volume effects, and we also find these to be roughly independent of the temperature up to our largest $T=189\textmd{ MeV}$. However, the signal-to-noise ratio becomes worse at high temperatures (which, in our fixed $N_t$ approach, correspond to finer lattices), again due to the cancellation of $\mathcal{O}(a^{-4})$
divergences in $\dena$ and $\dbna$. 

The fermionic anisotropies defined in \eq(\ref{eq def delta ferm}) also develop nonzero expectation values for $B>0$, see the right panel of \fig\ref{fig anisotropies one} for our zero temperature results for the up quark. We find the anisotropies $A(\F_f)$ to be negative for all three quark flavors. The anisotropy in physical units is by about a factor of five larger than the anisotropies found in the gluonic sector. 
Similarly as for the gluonic anisotropies, we find the magnitude of $A(\F_f)$ to be roughly independent of the temperature. 
We stress again that the anisotropies presented here are still subject to an additive renormalization, which we discuss in \secref{sec isoaniso} below. 

\subsection{Topological charge}

\begin{figure}[ht!]
\centering
 \includegraphics[width=0.49\linewidth]{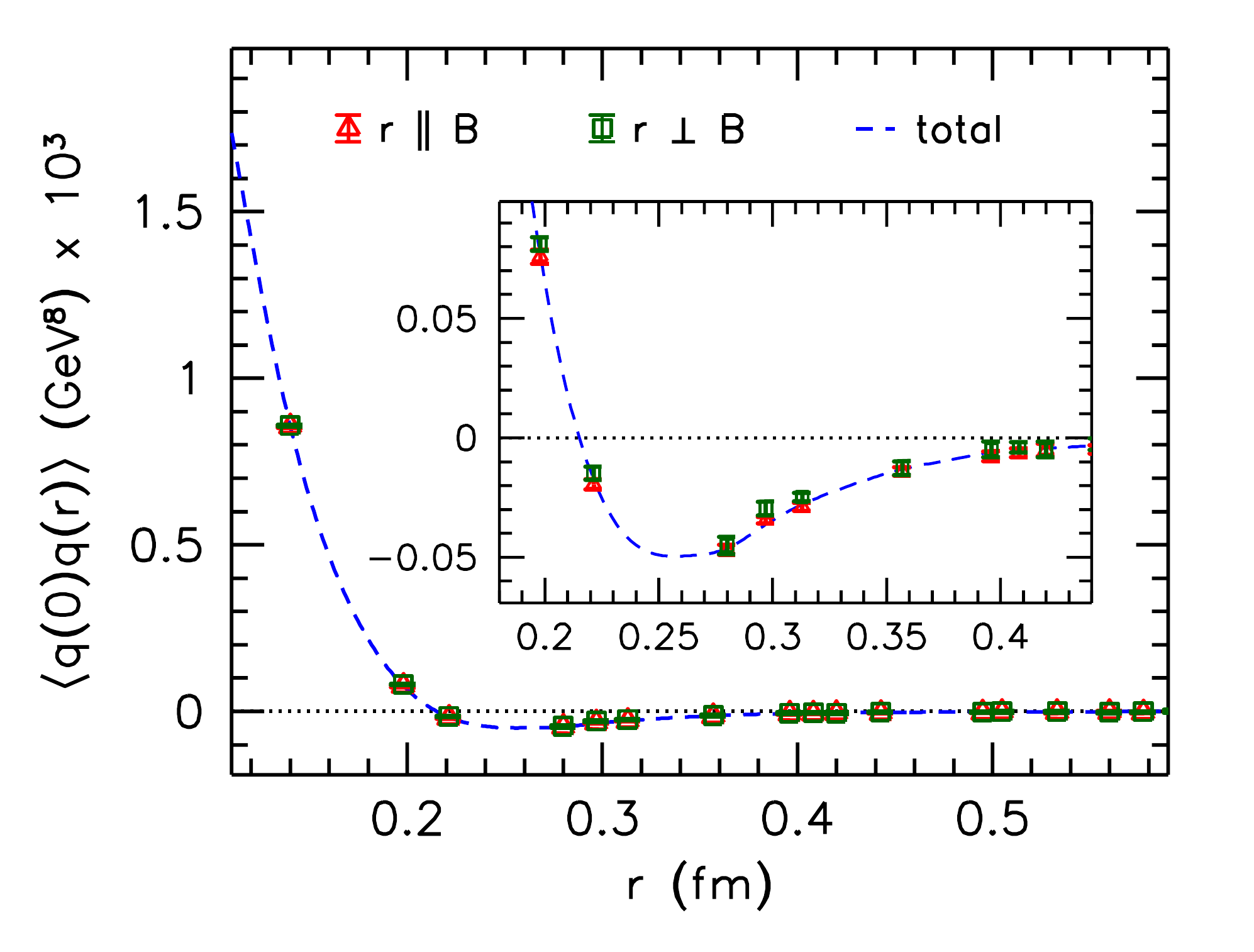}
 \includegraphics[width=0.49\linewidth]{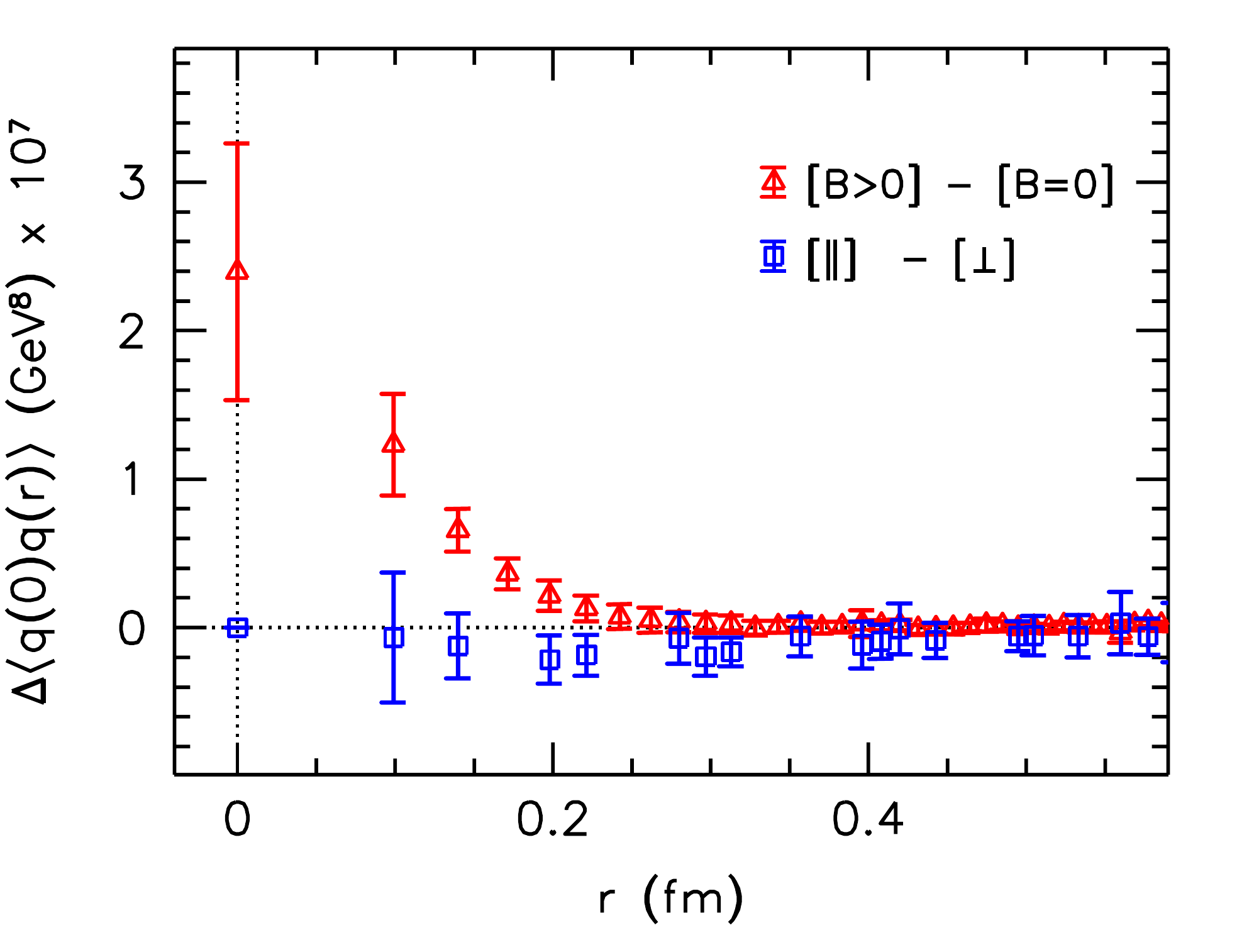}
\caption{Left panel: correlator of the topological charge density, \eq(\protect\ref{eq_top_corr_def}), at $e\ba=1.1\textmd{ GeV}^2$ at vanishing temperature on a $40^3\times 48$ lattice of lattice spacing $a=0.1 \textmd{ fm}$. The perpendicular and parallel (green and red points, respectively) correlators are compared to the total one (dashed line). Right panel: the difference between the total correlator at $e\ba=1.1\textmd{ GeV}^2$ and at $eB=0$ (red triangles), and the anisotropy between the parallel and perpendicular correlators at $e\ba=1.1\textmd{ GeV}^2$ (blue squares). }
\label{fig top}
\end{figure}

We proceed by investigating the correlator, \eq(\ref{eq_top_corr_def}), of the topological charge density. In the left panel of \fig\ref{fig top}, the parallel, perpendicular and total correlators are shown for our largest magnetic field, $e\ba=1.1\textmd{ GeV}^2$ on a $40^3\times 48$ lattice with $a=0.1\textmd{ fm}$, at $T=0$. The results exhibit the typical dependence of the correlator on the distance, with a positive peak at $r=0$, a zero at around $r=2 a$, and a negative tail for large distances. To quantify the effect of the magnetic field, in the right panel of the figure, we plot the difference between the parallel and perpendicular correlators (squares), and that between the $B>0$ and $B=0$ correlators (triangles). The latter reveals an enhancement of the correlator for small distances as $B$ grows. 
Note that due to the presence of the contact term, the correlator at $r=0$ diverges in the continuum limit (approximately as $a^{-6}$, as observed in ref.~\cite{Bruckmann:2011ve}). For fine lattices, this divergence dominates the difference created by the magnetic field, resulting in a lower signal/noise ratio for $r\lesssim 2a$. Altogether, we conclude that the magnetic field generates an $\mathcal{O}(10^{-7} \textmd{ GeV}^8)$ upward shift in the correlator at $r=0$.

We remark moreover that there is a slight tendency of a suppression of the parallel correlator with respect to the perpendicular one around $r=0.2-0.3\textmd{ fm}$, which is however, at the present level of our statistics, consistent with zero. 
Thus -- in sharp contrast to the pronounced effect of $B$ on the gauge action and on its components -- we conclude that the magnetic field induces no strong anisotropy in the topological charge density correlator, as was also conjectured in ref.~\cite{Basar:2011by}.

\section{Results II: Isotropic and anisotropic pressures}
\label{sec isoaniso}
We are now in the position to
discuss the isotropy properties of the pressure for nonzero magnetic fields. 
It is well known that the pressures are related to the spatial diagonal components of the energy-momentum tensor $T^{\alpha\beta}$~\cite{landau1971classical}. In the presence of external magnetic fields, $T^{\alpha\beta}$ has been studied extensively in the last decades. In several studies, the pressures obtained from these components were found to be anisotropic (as, e.g., in refs.~\cite{Canuto:1969ct,Martinez:2003dz,Ferrer:2010wz,Strickland:2012vu}). Such a difference between the longitudinal and transverse pressures would have important consequences in, e.g., the physics of magnetized neutron stars~\cite{Martinez:2003dz}. 
However, in the above studies, the currents generating the magnetic field are not taken into account, and thus, the considered system is open, with an energy-momentum tensor that is not conserved~\cite{jackson1975classical,rohrlich2007classical}. If the system is defined to include these currents as well, additional contributions arise~\cite{0022-3719-15-30-017,Potekhin:2011eb}, which are the topic of an ongoing discussion~\cite{Ferrer:2011ze,Potekhin:2011eb}.
Here, we address the problem from a somewhat different aspect, and consider the pressures as the response of the thermodynamic potential of the system against compressions in the corresponding directions.

\subsection{In the continuum}

Let us consider a polarizable medium in a finite volume $V=L_xL_yL_z$, at temperature $T=1/L_t$, exposed to a magnetic field $eB$ in the positive $z$ direction. The magnetic flux through the $x-y$ plane is given as $\Phi=eB\cdot L_xL_y$. The partition function of the system depends on the spatial extents, the temperature and the magnetic field, $\log\Z(L_i,T,eB)$, with $i=x,y,z$. 
The energy density $\epsilon$ and the pressures $p_i$ are given as total derivatives with respect to the corresponding extents,
\be
\epsilon=-\frac{1}{V} \frac{\d \log\Z}{\d (1/T)},\quad\quad\quad
p_i=\frac{T}{V} L_i\frac{\d \log\Z}{\d L_i}.
\ee
The reaction of the system to the simultaneous rescaling of all extents $L_\mu$, is related to the interaction measure (trace anomaly) $I$, which we have already defined in \eq(\ref{eq:tradef}). In terms of the energy density and the pressures it reads $I = \epsilon-p_x-p_y-p_z$.
The $B$-dependence of $\log\Z$ is related to the magnetization,
\be
M=\frac{T}{V} \frac{1}{e}\frac{\partial \log\Z}{\partial B}.
\label{eq:Mdef}
\ee

It is crucial now to distinguish between two distinct settings, which can be used to define the transverse pressures ($p_x$ and $p_y$) and the interaction measure. For the derivatives with respect to $L_x$ and $L_y$, one can keep fixed either the magnetic field ($eB=\textmd{const.}$), or the magnetic flux ($\Phi=\textmd{const.}$). We will refer to these cases as the $B$-scheme and the
$\Phi$-scheme, respectively. In the former case the flux grows as $L_x$ (or $L_y$) is increased, whereas in the latter case, $eB$ decreases for increasing $L_x$ (or $L_y$). 
Along constant $eB$, the only dependence of $\log\Z$ on the extents is the explicit one, whereas for the $\Phi$-scheme, an additional term emerges due to the implicit dependence of $B$ on $L_x$ and $L_y$,
\be
p_i^{(B)} = \frac{T}{V} \frac{\partial \log\Z}{\partial \log L_i}, \quad\quad\quad
p_i^{(\Phi)} = \frac{T}{V} \frac{\partial \log\Z}{\partial \log L_i} + \frac{T}{V} \frac{\partial \log\Z}{\partial B} \cdot \left.\frac{\partial B}{\partial \log L_i}\right|_{\Phi},
\label{eq:presisoaniso}
\ee
revealing the relation between the pressures in the two schemes. The energy density -- just as $p_z$ -- is independent of the scheme,
\be
p_z^{(\Phi)}=p_z^{(B)}, \quad\quad\quad
p_{x,y}^{(\Phi)}=p_{x,y}^{(B)}-M\cdot eB, \quad\quad\quad
\epsilon^{(\Phi)} = \epsilon^{(B)},
\label{eq:BorPhip}
\ee
which makes the interaction measure also scheme-dependent,
\be
I^{(\Phi)} = I^{(B)}+2M\cdot eB.
\label{eq:IPHIB}
\ee
Below we will omit the superscripts from the scheme-independent
quantities $\epsilon$ and $p_z$. Note that at zero temperature, the $z$ and $t$ directions are indistinguishable, therefore here $I=-2(p_x+p_z)$, implying the same scheme-dependence as \eq(\ref{eq:IPHIB}).

Let us now consider a homogeneous system exposed to the external field, with the thermodynamic potential proportional to the three-volume, $T\log\Z(L_i,T,eB)= V\cdot \Omega(eB,T)$. Such a homogeneous thermodynamic potential describes, for example, free charged particles (see, e.g., ref.~\cite{Endrodi:2013cs}). On the
one hand, the volume only appears as the proportionality factor so
that in the $B$-scheme, homogeneity implies {\it isotropic} pressures. On the other hand, expressed as function of the magnetic flux, $T\log\Z(L_i,T,\Phi)=V\cdot \Omega(\Phi/(L_xL_y),T)$, which is not proportional to the volume anymore. Thus, in the $\Phi$-scheme, one obtains {\it anisotropic} pressures. The difference between the pressures in the two schemes is fixed by \eq(\ref{eq:BorPhip}):
\be
p_{x,y}^{(B)}=p_z, \quad\quad\quad p_{x,y}^{(\Phi)}=p_z-M\cdot eB.
\label{eq:1}
\ee
The two schemes give different transverse pressures, since they result from a different compression of the system: either pushing the magnetic field lines together with the box in which the system resides ($\Phi$-scheme), or leaving the field lines unaffected ($B$-scheme). In the $\Phi$-scheme, one may therefore think of the field lines as being frozen in. This, for instance, is generally assumed to be the case for the magnetohydrodynamics of a perfectly conducting plasma~\cite{Vachaspati:1991nm,Kandus:2010nw}.
The longitudinal pressure $p_z$ is however independent of the scheme, since both $B$ and $\Phi$ remain constant during the compression of the system in the direction of the magnetic field.

One also has to specify the scheme when defining the pressures as the spatial diagonal components of the energy-momentum tensor. The tensor $T_{\alpha\beta}$ can be obtained by considering the variation of the action with respect to the metric $g^{\alpha\beta}$. Again, such variations can be performed keeping either $B$ or $\Phi$ fixed. The conventional definition of the energy-momentum tensor (as used in, e.g., ref.~\cite{Ferrer:2010wz}) corresponds to the $\Phi$-scheme:
$\Phi$ as a scalar quantity is not affected by general coordinate transformations, while fixing the two-form $eB$ breaks general covariance. 
This can also be seen by considering the covariant derivative in a transversal direction, $D_{x,y}=\partial_{x,y} \pm i eB r_{y,x}$. Rescaling the coordinates in the $x$-$y$ plane, $(x,y)\mapsto(\xi x, \xi y)$, $(\partial_x, \partial_y) \mapsto (\xi^{-1}\partial_x, \xi^{-1}\partial_y)$, the two terms in the covariant derivative only transform in the same way (i.e., gauge covariance is only maintained), if $B$ transforms as $B\mapsto \xi^{-2}B$. This indeed corresponds to $\Phi\mapsto\Phi$. Accordingly, the anisotropy, \eq(\ref{eq:1}), between the pressures in the $\Phi$-scheme agrees with the anisotropy defined through $T_{\alpha\beta}^{(\Phi)}$~\cite{Ferrer:2010wz}, with $I^{(\Phi)} =\trace T^{(\Phi)}$ (and similarly $I^{(B)}=\trace T^{(B)}$).

We remark that in general, besides the magnetization of the polarizable medium, $\log\Z$ also contains the energy of the magnetic field itself, $-V/T\cdot B^2/2$. (Note that in this term the magnetic field appears separately and not in the product with the charge.) Therefore, in the derivative with respect to $B$, an additional term $-B$ appears, which we have neglected above. The difference of the magnetic field $B$ and the magnetization equals the external magnetic field $H$, which would have been present in the absence of the medium~\cite{landau1995electrodynamics}, $H=B-Me$. If this static magnetic energy of the background field is also taken into account, $M\cdot eB$ has to be replaced by $(Me-B)B=-HB$ in \eqs(\ref{eq:BorPhip})--(\ref{eq:1}).

\subsection{On the lattice}

Let us now consider the lattice setup with extents $L_\mu=N_\mu a$ and lattice spacing $a$. The natural variable describing the magnetic field on the lattice is the flux $\Phi$, which is quantized, due to the finiteness of the lattice volume in the directions perpendicular to the magnetic field,
\be
\Phi = q_dB \cdot N_s^2 a^2 = 2\pi N_b,\quad\quad\quad N_b\in\mathds{Z},
\label{eq:quant}
\ee
where the down quark charge $q_d$ enters. 
Therefore, the lattice system automatically realizes the $\Phi$-scheme, since the flux cannot be changed continuously. In accordance with \eq(\ref{eq:presisoaniso}), we then expect the lattice pressures to develop an anisotropy for nonzero $B$. Furthermore, we neglect the pure magnetic energy $-B^2/2$ in $\log\Z$, since it only constitutes a constant irrelevant shift in the path integral of \eq(\ref{eq:partfunc}), and cancels from all expectation values. This implies that the magnetic field only enters $\log\Z$ in the renormalization group invariant combination $q_fB\sim eB$.
Altogether, the lattice pressures therefore satisfy
\be
p_x-p_z=p_y-p_z=-M\cdot eB.
\label{eq:latpres}
\ee

\begin{figure}[ht!]
\centering
 \includegraphics[width=0.49\linewidth]{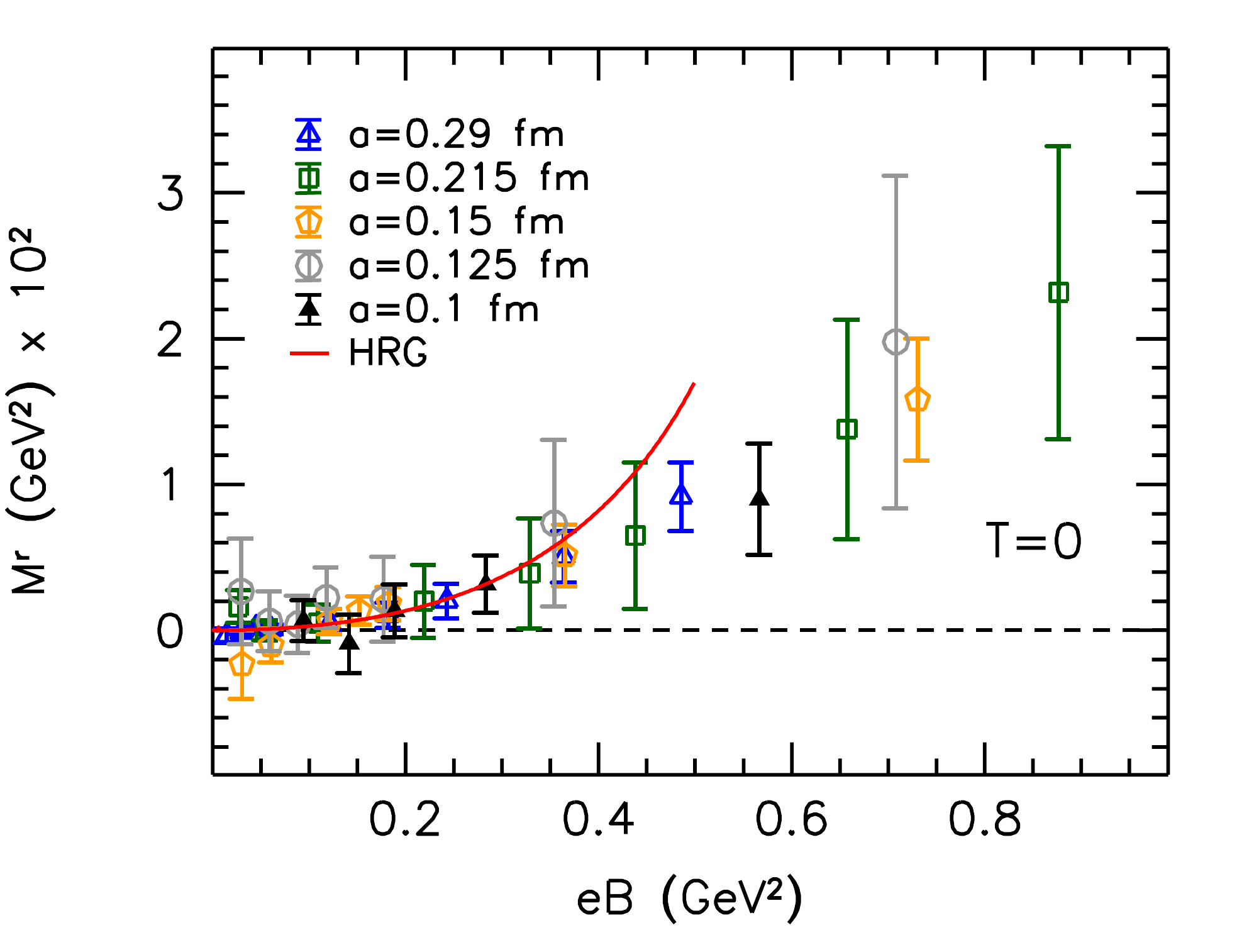}
\caption{Estimate of the renormalized magnetization of QCD. The positivity of $M^r$ indicates that the QCD vacuum is paramagnetic. Shown are lattice results on five different lattice spacings (colored points) and the hadron resonance gas curve~\protect\cite{Endrodi:2013cs} (solid line).}
\label{fig magnetization}
\end{figure}

Due to the quantization, \eq(\ref{eq:quant}), the magnetization cannot be determined on the lattice as the partial derivative of $\log \Z$ with respect to the magnetic field. However, we can make use of the anisotropy \eq(\ref{eq:latpres}) of the $\Phi$-scheme pressures, and compute the magnetization $M$ from the difference between the transverse and longitudinal components. As shown in \app\ref{sec pressure}, the anisotropies of both the gauge and fermionic actions enter this difference,
\be
  -M\cdot eB= -(\zeta_g+\hat{\zeta}_g) \cdot\left[A(\bna)-A(\ena)\right] - \zeta_f \cdot \sum_f A(\F_f),
\label{eq:openaniso}
\ee
where $\zeta_g$, $\hat\zeta_g$ and $\zeta_f$ are renormalization coefficients for the anisotropic lattice parameters, which are of the form $1+\mathcal{O}(g^2)$ in perturbation theory~\cite{Karsch:1982ve,Karsch:1989fu}, and are expected to be of $\mathcal{O}(1)$ for the couplings used in the present study. 
We have seen in \secref{sec:anisotropies} that $\sum_f A(\F_f)$ is by about a factor of five larger than $A(\bna)-A(\ena)$, and therefore dominates the right hand side of \eq(\ref{eq:openaniso}). Hence, for a first estimate, we neglect the gluonic term and also assume $\zeta_f=1$. The final step is to perform the additive renormalization of $M\cdot eB$, which contains the logarithmic divergence $\sim(eB)^2\log a$. As discussed in \app\ref{sec renorm}, this renormalization at $T=0$ corresponds to subtracting the {\it total} $\mathcal{O}((eB)^2)$ contribution in $M\cdot eB$. This is done by fitting the magnetization at small magnetic fields and extrapolating the result to zero $eB$:
\be
M^r \cdot eB = M\cdot eB - (eB)^2\cdot \lim_{eB\to0} \frac{M\cdot eB}{(eB)^2}.
\label{eq:Mrenorm}
\ee
The estimated renormalized magnetization is plotted in \fig\ref{fig magnetization}, as a function of $eB$. We observe that after subtracting the $\mathcal{O}((eB)^2)$ contribution, the magnetization becomes positive, indicating a {\it paramagnetic} QCD vacuum. The results for $M^r$ are also compared here to the hadron resonance gas (HRG) model prediction~\cite{Endrodi:2013cs}, showing a remarkable agreement for $eB<0.4\textmd{ GeV}^2$. Taking into account the independence of $A(\F_f)$ on the temperature, as discussed in subsec.~\ref{sec:anisotropies}, we conclude that the magnetization remains roughly constant up to the transition region. This is also consistent with the HRG description of ref.~\cite{Endrodi:2013cs}. Note that
in ref.~\cite{Bali:2012jv} we presented results 
on the magnetic susceptibility of the quark spins,
which indicated a diamagnetic behavior of this contribution. Here,
we addressed the total magnetization of the QCD vacuum, which -- besides the spin term -- also includes
the quark angular momentum contributions. 

Let us summarize this section. We have shown that the pressure in the presence of an external magnetic field is isotropic if the $B$-scheme is applied, i.e., if the magnetic field $eB$ is kept constant while compressing the system. Conversely, keeping the magnetic flux fixed ($\Phi$-scheme) results in anisotropic pressures, and the difference between $p_x$ and $p_z$ turns out to be proportional to the magnetization $M$. Due to the quantization of the magnetic flux, $M$ is not directly accessible on the lattice. However, exploiting the fact that the lattice implementation corresponds to the $\Phi$-scheme, we can express $M$ as an expectation value of anisotropies, which can be computed directly on the lattice.
The only complication is due to the anisotropy coefficients $\zeta_g$, $\hat\zeta_g$ and $\zeta_f$, which should ideally be determined non-perturbatively.
However, a first estimate of $M^r$ can be obtained by neglecting the gluonic contribution, and assuming $\zeta_f=1$.
The full determination of the magnetization, involving the calculation of these coefficients, will be performed in a forthcoming study.

\section{Conclusions}
\label{sec conclusions}

In this paper, we have presented a comprehensive analysis of the response of gluonic observables of the QCD vacuum to an external (Abelian) magnetic field. We performed lattice simulations using dynamical staggered quarks with physical masses at several lattice spacings, temperatures and external field strengths. 
As gluons themselves do not carry electric charge, they are only affected by $B$ indirectly, through their coupling to electrically charged quarks. 
Still, we found several gluonic quantities that show a pronounced dependence on $B$. 

First, the gluonic contribution to the interaction measure was determined, which we observed to undergo magnetic catalysis at low temperatures, and inverse magnetic catalysis above the transition temperature. It is instructive to compare this observable to the fermionic contributions (which we have shown to be given by the quark condensates), exhibiting the same qualitative tendencies. This reflects the distinct correlation between the reaction of the gluonic and fermionic sectors to the external magnetic field, through the strong QCD coupling.

Second, we found the external field to induce an anisotropy between the individual components of the squared field strengths, separating the chromo-electric and chromo-magnetic fields in the directions parallel and perpendicular to $B$. A similar tendency is visible in the fermionic sector, where the anisotropy was found to be by about a factor of five larger. We have shown that the sum of these anisotropies (up to $\mathcal{O}(1)$ factors related to lattice anisotropy renormalization) equals the QCD magnetization $M$ -- an observable which is otherwise not directly accessible by standard lattice methods. After performing the additive renormalization of $M$ through charge renormalization, we estimated the magnetization by assuming trivial anisotropy factors. We have obtained $M>0$, which indicates that the response of the QCD vacuum is {\it paramagnetic}. We remark that paramagnetic quarks have been shown to imply a decreasing transition temperature based on large $N_c$ arguments~\cite{Fraga:2012ev}, in line with the lattice results~\cite{Bali:2011qj,Bali:2012zg}.

Third, we determined the correlator of the topological charge density for our largest magnetic field $eB=1.1\textmd{ GeV}^2$. 
In contrast to the above results, we found no significant difference between the correlators in the directions parallel and perpendicular to $B$. At the same time, our findings indicate an $\mathcal{O}(10^{-7} \textmd{ GeV}^8)$ upward shift of the correlator at small distances caused by the magnetic field. 

To obtain qualitative insight into the response of gluonic observables to the external field, we examined a combined Euler-Heisenberg effective action for QED and QCD. This analysis shows that there is, on the one hand, no coupling of the external field to terms which contribute to the topological charge density and, on the other hand, that there is an anisotropic coupling to the electric and magnetic components of the gluon field strength tensor, with a tendency, which is in accordance with our numerical results.

\acknowledgments

We thank the ECT$^\star$ in Trento for their hospitality during the workshop ``QCD in strong magnetic fields'', where various ideas related to the present paper were discussed. Moreover, we would like to thank Szabolcs Bors\'anyi, Pavel Buividovich, Eduardo Fraga, S\'andor Katz and K\'alm\'an Szab\'o for useful discussions.
This work has been supported by the EU Initial Training Network STRONGnet, by the DFG under grants SFB/TR-55 and BR 2872/4-2 and by the Alexander von Humboldt foundation. FG acknowledges support by the ``Bayerische Elitef\"orderung''.

\appendix

\section{Renormalization at nonzero \texorpdfstring{\boldmath{$B$}}{B}}
\label{sec renorm}

In this appendix, we discuss the renormalization of our observables at a nonzero external magnetic field. In general, the logarithm of the partition function contains additive divergences in the cutoff $\Lambda$, which, on the lattice, corresponds to the inverse lattice spacing $\Lambda\sim a^{-1}$. These divergences include quartic, quadratic and logarithmic terms at $B=0$~\cite{Leutwyler:1992yt}, whereas for nonzero magnetic fields, an additional logarithmic divergence appears, which is canceled by electric charge renormalization~\cite{Schwinger:1951nm,Elmfors:1993bm}, see also refs.~\cite{Dunne:2004nc,Endrodi:2013cs}. Thus, in a finite four-volume of size $V_4=T/V$, $\log\Z$ has the divergence structure
\be
\log\Z = c_1 \cdot V_4a^{-4} + c_2 \cdot V_4m_f^2a^{-2} + c_3 \cdot V_4m_f^4\log a + c_4\cdot \Phi^2\log a
+ \textmd{finite}.
\label{eq:divs}
\ee
We remark that the factor accompanying the $\Phi$-dependent divergence is related to the leading coefficient of the $\beta$-function of the theory~\cite{Dunne:2004nc,Endrodi:2013cs}, $c_4\cdot \Phi^2 \log a \sim \beta_1V_4  \cdot (eB)^2 \log a$, and the proportionality factor is given by the squared quark charges $\sum_f (q_f/e)^2$, which we suppress in the following.

Clearly, the first three divergences in \eq(\ref{eq:divs}) cancel when computing differences $\Delta$ in the magnetic field, see \eq(\ref{eq:delta}). Since the $\Phi$-dependent divergence is $m_f$-independent, one easily notices that all divergences cancel from the difference of the condensate $m_f \Delta\bar\psi_f\psi_f$. Moreover, the derivative of the magnetic field-dependent divergence with respect to $\log a$ at constant $\Phi$ is already finite, making the interaction measure \eq(\ref{eq:tradef}) in the $\Phi$-scheme, $\Delta I^{(\Phi)}$, also completely free of divergences. Note however, that the magnetic field-dependent divergence does not cancel if $eB\sim\Phi/a^2$ is kept constant, and thus $\Delta I^{(B)}$ still contains a logarithmic divergence. The magnetization of \eq(\ref{eq:Mdef}) also inherits the logarithmic divergence. Expressing these divergences as functions of $eB$, we obtain
\be
M\cdot eB = 2 \beta_1 \cdot (eB)^2\log a+\textmd{finite},\quad\quad\;
\Delta I^{(\Phi)} = \textmd{finite},\quad\quad\;
\Delta I^{(B)} = -4 \beta_1 \cdot (eB)^2\log a + \textmd{finite},
\label{eq:2}
\ee
which is in accordance with \eq(\ref{eq:IPHIB}). 
For the sake of completeness, we also give the divergences of the energy density and the pressures. In the $B$-scheme, the pressures are proportional to $\log\Z$. Thus,
\be
p_{x,y}^{(B)}=p_z = \beta_1 \cdot (eB)^2\log a + \textmd{finite}.
\label{eq:3}
\ee
Using \eqs(\ref{eq:2}),~(\ref{eq:3}) and (\ref{eq:1}), the energy density and the pressures in the $\Phi$-scheme then read
\be
\epsilon= -\beta_1\cdot (eB)^2\log a + \textmd{finite},\quad\quad\;
p_{x,y}^{(\Phi)} = -\beta_1\cdot (eB)^2\log a + \textmd{finite}, \quad\quad\;
p_{z} = \beta_1\cdot (eB)^2\log a + \textmd{finite}.
\label{eq:4}
\ee

These logarithmic divergences are eliminated by the simultaneous renormalization of the electric charge and of the bare magnetic field~\cite{Schwinger:1951nm,Elmfors:1993bm,Dunne:2004nc,Endrodi:2013cs},
\be
\frac{B^2}{2} - \beta_1 (eB)^2 \log a = \frac{B_r^2}{2}.
\ee
Since we are not interested in the static magnetic background energy $B_r^2/2$, at zero temperature this renormalization corresponds to subtracting the total $\mathcal{O}((eB)^2)$ contribution from $\log\Z$ -- and, thus, from the observables of \eqs(\ref{eq:2})--(\ref{eq:4}). In particular, for the magnetization this renormalization amounts to the procedure carried out in \eq(\ref{eq:Mrenorm}).

\section{Improvement of the interaction measure}
\label{sec imp}

In this appendix, we discuss the scaling properties of the interaction measure (trace anomaly) $I$.
Using its definition \eq(\ref{eq:tradef}), $I$ can be written in terms of partial derivatives with respect to the lattice parameters $\beta$, $am_f$ and $\Phi$. 
However, to measure $I$ in the $\Phi$-scheme of \app\ref{sec isoaniso}, the magnetic flux is to be kept constant, therefore the partial derivative with respect to $\Phi$ does not contribute.
Altogether, $I$ can be decomposed into gluonic and fermionic contributions
\be
I = -\frac{T}{V}\frac{\d \log\Z}{\d \log a} = I_g + \sum_f I_f, \quad\quad
I_g = -\frac{\partial \beta}{\partial \log a} \cdot \frac{T}{V}\frac{\partial \log\Z}{\partial \beta},\quad\quad
I_f =-\frac{\partial (am_f)}{\partial \log a} \cdot \frac{T}{V}\frac{\partial \log\Z}{\partial (am_f)},
\label{eq_pre}
\ee
such that the change in these contributions due to the magnetic field reads
\be
\Delta I_g = - R_\beta \cdot\Delta s_g, \quad\quad\quad
\Delta I_f = -R_{m_f} \cdot m_f \Delta \bar{\psi}_f\psi_f,
\label{eq:Rmf}
\ee
where we used the definition of the gluonic action density and the quark condensate densities, \eq(\ref{eq_the_quantities}), and define the scaling factors
(that are related to the lattice $\beta$- and $\gamma$-functions)
\be
R_\beta = -\frac{\partial \beta}{\partial \log a}, \quad\quad\quad 
R_{m_f} = \frac{\partial \log(am_f)}{\partial \log a},
\label{eq:Rdefs}
\ee
which have been determined non-perturbatively, along the line of constant physics (LCP) for the action we use in ref.~\cite{Borsanyi:2010cj}. 
We remark that the external magnetic field affects neither the lattice spacing, nor the LCP (see discussion in ref.~\cite{Bali:2011qj}), therefore $R_\beta$ and $R_{m_f}$ are independent of $\Phi$.
We have shown in \app\ref{sec renorm} that both the change in the condensate $m_f\Delta \bar\psi_f\psi_f$ and the change in the interaction measure $\Delta I=\Delta I^{(\Phi)}$ (the sum of the two terms of \eq(\ref{eq:Rmf})) are finite in the continuum limit. We remark moreover that both quantities are expected to approach the continuum limit as $\mathcal{O}(a^2)$, in line with the scaling properties of the employed action.

The factor $R_{m_f}$ in front of the condensate in \eq(\ref{eq:Rmf}) is related to the lattice $\gamma$-function, cf. \eq(19) in ref.~\cite{Cheng:2007jq},
\be
R_{m_f}=1+\gamma_f(a),\quad\quad\quad \gamma_f=\mathcal{O}(g^2) = \mathcal{O}\left(1/\log a\right),
\label{eq:rmfvanish}
\ee
where the anomalous dimension $\gamma_f$ is inherited from the mass renormalization factor $Z_m$ \cite{Lee:1993af}. This shows that $\Delta I_f$ (and then also $\Delta I_g=\Delta I - \sum_f\Delta I_f$) is also separately finite in the limit $a\to0$. However, in contrast to the total $\Delta I$, the logarithmic convergence of $R_{m_f}$ makes $\Delta I_f$ and $\Delta I_g$ converge slowly towards the continuum limit. 

\begin{figure}[ht!]
\centering
 \includegraphics[width=0.5\linewidth]{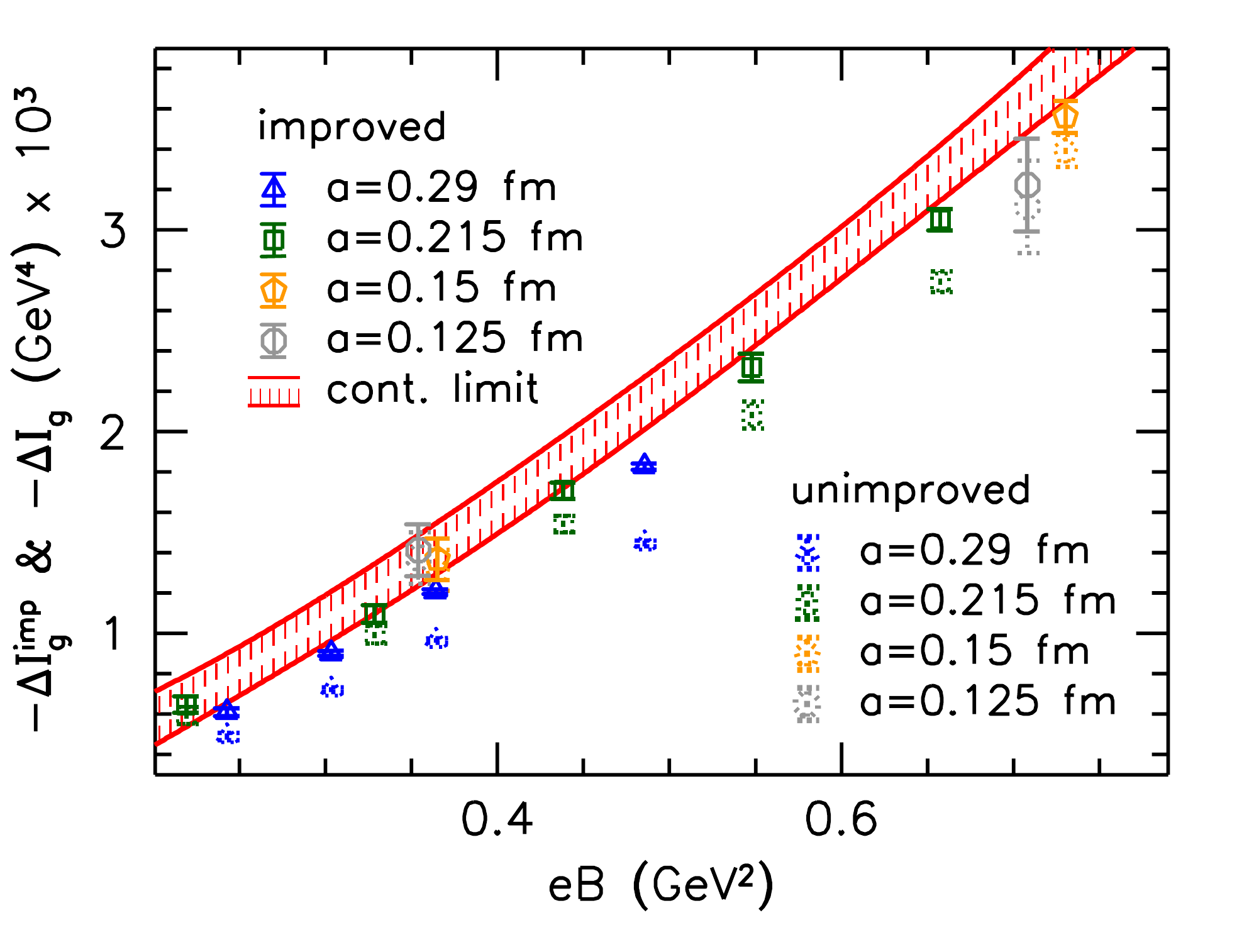}
\caption{Visualization of the effect of the improvement of the gluonic contribution to the interaction measure, \eq(\ref{eq_post}): the $\mathcal{O}(1/\beta)=
\mathcal{O}(1/\log a)$ scaling is improved to $\mathcal{O}(a^2)$. The red band is the continuum limit based on the improved data (solid points), as in the left panel of \fig\ref{fig gluonic}, whereas dotted points correspond to the unimproved data.
}
\label{fig impr}
\end{figure}

In order to improve this slow convergence, we regroup the fermionic and gluonic contributions in $\Delta I$ in the following way,
\be
\Delta I = \Delta I_g + \sum_f \Delta I_f \equiv \Delta I_g^{\rm imp} + \sum_f \Delta I_f^{\rm imp},
\ee
\be
\Delta I_g^{\rm imp} = -R_\beta \cdot \Delta s_g -\sum_f \left( R_{m_f} - 1\right) \cdot m_f\Delta\bar{\psi}_f\psi_f, \quad\quad\quad
\Delta I_f^{\rm imp} = -\sum_f m_f \Delta \bar\psi_f\psi_f,
\label{eq_post}
\ee
recovering \eq(\ref{eq_the_long_name_quantity}) for the improved gluonic contribution to $\Delta I$. 
Note that the improvement correction is proportional to $R_{m_f}-1$, which vanishes in the continuum limit, in accordance with \eq(\ref{eq:rmfvanish}). Furthermore, \eq(\ref{eq_post}) shows that $-\Delta I_f^{\rm imp}$ is given just by the renormalized quark condensate (which we have studied in detail in ref.~\cite{Bali:2012zg}), implying that its convergence is improved to $\mathcal{O}(a^2)$. Therefore, $\Delta I_g^{\rm imp}$ is also expected to scale as $\mathcal{O}(a^2)$. The effect of this non-perturbative improvement is shown in \fig\ref{fig impr}.

\section{Details on the pressure anisotropy and the magnetization}
\label{sec pressure}

In this appendix, we derive \eq(\ref{eq:openaniso}), and show how the gluonic and fermionic anisotropies are related to the difference of the transverse and longitudinal pressures in the $\Phi$-scheme realized on the lattice. Let us write down again the functional integral in Euclidean space-time, as in \eq(\ref{eq:partfunc}),
\be
\Z = \int \D U e^{-\beta S_g} \, \prod_f \det\left[a\slashed{D} + am_f\right],
\label{eq:pf}
\ee
where the product runs over quark flavors $f$. To relate the energy density and the pressures directly to the partition function~\eq(\ref{eq:pf}), we need to consider anisotropic lattice spacings~\cite{Engels:1981qx,Karsch:1982ve}. We define a lattice spacing $a_{\alpha}$ in direction $\alpha$ while $a=\xi a_{\alpha}$ denotes the lattice spacing in the remaining three space-time directions. The physical anisotropy -- which can be determined, e.g., from the static potential~\cite{Bali:2000un} or using the Wilson flow~\cite{Borsanyi:2012zr} -- is given by
\be
\xi= \frac{a}{a_\alpha}.
\ee
This physical anisotropy differs from the bare anisotropies $\xi_{g0}$ and $\xi_{f0}$, which enter the gauge and fermionic actions, respectively, as parameters. In the gauge sector, the total action $S_g$ -- as in \eq(\ref{eq:sgsym}) -- splits into two parts,
\be
S_g = \xi_{g0} \cdot P_{\alpha} + \frac{1}{\xi_{g0}} \cdot \hat{P}_{\alpha}, \quad\quad\quad
P_{\alpha}\equiv\sum_{\mu\neq\alpha} \sum_n \frac{1}{3}\Re\trace P_{\mu \alpha}(n), \quad\quad
\hat{P}_{\alpha}\equiv\sum_{\substack{\mu,\nu\neq \alpha\\\mu<\nu}} \sum_n \frac{1}{3} \Re\trace P_{\mu\nu}(n),
\label{eq:pphat}
\ee
with $P_{\mu\nu}(n)$ being the Symanzik tree-level improved plaquettes~\cite{Weisz:1982zw},
\be
\label{eq:weisz}
P_{\mu\nu}(n) = -\frac{1}{12} \left(\mathds{1}-U_{\mu\nu}^{2\times 1}(n)\right) + \frac{5}{3} \left(\mathds{1}-U_{\mu\nu}^{1\times 1}(n)\right),
\ee
where $U_{\mu\nu}^{(m,n)}$ denotes the product of gauge links along
the boundaries of planar rectangles of size $m\times n$ lattice
units in the $\mu$-$\nu$ plane.
Note that two separate couplings ($\beta$ and $\xi_{g0}(\beta)$) need to be introduced in the gluonic sector to define the continuum limit along a line of fixed physical
$\xi$~\cite{Karsch:1982ve}. 
In the fermionic sector, the anisotropy shows up as an extra factor of $\xi_{f0}$ in the $\alpha$ direction in the Dirac operator, see, e.g., refs.~\cite{Karsch:1989fu,Levkova:2006gn}. Schematically, $\slashed{D}$ reads
\be
a\slashed{D} = \xi_{f0} \cdot a\slashed{D}_\alpha + \sum_{\mu\neq\alpha} a\slashed{D}_\mu,\quad\quad\quad
a\slashed{D}_\mu = \frac{1}{2} \left(\eta_\mu U_\mu u_\mu- \eta_\mu U_\mu^\dagger u_\mu^*\right),
\label{eq:diracop}
\ee
with $\eta_\mu$ being the staggered phases, $U_\mu$ the $\mathrm{SU}(3)$ links and $u_\mu$ the $\mathrm{U}(1)$ factors, which generate the magnetic field $B$.
 In the end, we will set the anisotropy to $\xi_{g0}=\xi_{f0}=1$ (which results in $\xi=1$). However, to calculate individual components of the energy momentum tensor, $\xi_{f0}$ and $\xi_{g0}$ are to be treated as free parameters.

Now, to obtain the energy density, we consider $\alpha=t$,
\be
\epsilon = -\frac{1}{V}\left.\frac{\d \log\Z}{\d (1/T)}\right|_{a}
=
\xi^2\frac{1}{N_s^3N_t}\frac{1}{a^4}\left.\frac{\d \log\Z}{\d \xi}\right|_{a},
\label{eq:edef}
\ee
with $V=(N_sa)^3$ being the three-volume and $T=(N_ta_t)^{-1}$ the temperature. 
For the pressures we take $\alpha$ as spatial,
\be
p_\alpha = \frac{T}{V}L_\alpha\left.\frac{\d \log\Z}{\d L_\alpha}\right|_{a}
=
-\xi^2\frac{1}{N_s^3N_t}\frac{1}{a^4}\left.\frac{\d \log\Z}{\d \xi}\right|_{a},
\label{eq:pdef}
\ee
where $L_\alpha=N_s a_\alpha$ is the size of the lattice in the $\alpha$ direction and in this case $V=N_s^3a^2a_\alpha$ is the three-volume and $T=(N_ta)^{-1}$ the temperature. 

From \eqs(\ref{eq:pphat}) and~(\ref{eq:diracop}), one easily obtains for the derivative of the partition function~(\ref{eq:pf}) with respect to the physical anisotropy $\xi$
\be
\label{eq:dlogzdxi}
\begin{split}
\left.\frac{\d \log\Z}{\d \xi}\right|_{a} 
&=
\frac{\partial (\beta\xi_{g0})}{\partial \xi} \expv{-P_{\alpha}} + 
\frac{\partial (\beta/\xi_{g0})}{\partial \xi} \big\langle{-\hat{P}_{\alpha}}\big\rangle \\
&\hspace*{.5cm}+ 
\sum_f\frac{\partial (am_f)}{\partial \xi}\expv{\bar\Psi_f\Psi_f} +
\frac{\partial \xi_{f0}}{\partial \xi}\sum_f\left[ \expv{\bar\Psi_f \slashed{D}_\alpha \Psi_f}
- \frac{N_s^3N_tN_c}{4}\right] ,
\end{split}
\ee
where we used the definition \eq(\ref{eq_the_quantities}) of the condensates and the identities\footnote{We suppress Grassmannian integration on the right hand side of \eq(\ref{eq:aux1}), and also in \eq(\ref{eq:aux2}) below.},
\be
\frac{\partial}{\partial \xi_{f0}} \log \det (a\slashed{D}+am_f) =
\Trace \frac{a\slashed{D}_\alpha }{a\slashed{D}+am_f } \equiv \bar\Psi_f\slashed{D}_\alpha \Psi_f.
\label{eq:aux1}
\ee
The above trace is taken over space-time and color indices (note that in the staggered formulation, spinor indices are absent). 
The last term in \eq(\ref{eq:dlogzdxi}), proportional to the number of colors $N_c=3$, is chosen such that the energy density and the pressures according to \eqs(\ref{eq:edef}) and~(\ref{eq:pdef}) are consistent with the definition of the interaction measure, \eq(\ref{eq_pre}), see below.  
Note moreover that the parameters of $\log\Z$ -- besides $\beta$, $am_f$, $\xi_{g0}$ and $\xi_{f0}$ -- also include the magnetic flux $\Phi\sim a^2eB$ for nonzero magnetic fields. To obtain the pressures in the $\Phi$-scheme of \secref{sec isoaniso}, we have to consider the partial derivatives at fixed $\Phi$. Therefore, the partial derivative with respect to $\Phi$ does not contribute to \eq(\ref{eq:dlogzdxi}).

Let us now consider \eq(\ref{eq:dlogzdxi}) at $\xi=1$. The magnetic field does not affect the line of constant physics for the quark masses (see the discussion in ref.~\cite{Bali:2011qj}). Therefore, $\partial (am_f)/\partial \xi$ is independent of the direction $\alpha$, in which the anisotropy was introduced, and the corresponding term cancels from the difference between $p_x$ and $p_z$. Similarly, $B$ has no effect on the lattice spacing~\cite{Bali:2011qj}. Therefore, the derivatives of the inverse gauge couplings $\partial (\beta\xi_{g0})/\partial \xi$, $\partial (\beta/\xi_{g0})/\partial \xi$ and $\partial \xi_{f0}/\partial \xi$ are also independent of $\alpha$. Altogether, we obtain for the difference of the pressures,
\be
p_x-p_z = -\frac{T}{V} \bigg[ \zeta_g \beta \cdot\expv{-P_x+P_z} - 
\hat{\zeta}_g \beta \cdot \langle{-\hat{P}_x+\hat{P}_z}\rangle
+
\zeta_f \cdot \sum_f\expv{\bar\Psi_f \slashed{D}_x \Psi_f -\bar\Psi_f \slashed{D}_z \Psi_f}\bigg],
\ee
and an analogous expression for $p_y-p_z$. Above we introduced the
abbreviations
\be
\zeta_g \equiv \frac{1}{\beta}\left.\frac{\partial (\beta\xi_{g0})}{\partial \xi}\right|_{\xi=1},\quad\quad\quad
\hat{\zeta}_g \equiv -\frac{1}{\beta}\left.\frac{\partial (\beta/\xi_{g0})}{\partial \xi}\right|_{\xi=1},
\quad\quad\quad
\zeta_f \equiv \left.\frac{\partial \xi_{f0}}{\partial \xi}\right|_{\xi=1},
\ee
in a manner that in perturbation theory, both gluonic coefficients $\zeta_g$ and $\hat\zeta_g$~\cite{Karsch:1982ve}, and also the fermionic coefficient $\zeta_f$~\cite{Karsch:1989fu} are $1+\mathcal{O}(g^2)$.
Using \eq(\ref{eq:pphat}), and writing out explicitly the components of $P_{\mu\nu}$ according to \eq(\ref{eq:compgact}), we get,
\be
-P_\rho+P_z = \hat{P}_\rho-\hat{P}_z = \frac{1}{6}\sum_n \left[\trace\bna^2_\rho(n)-\trace\bna^2_z(n) -\trace\ena^2_\rho(n)+\trace\ena^2_z(n)\right],\quad\quad\quad \rho=x,y.
\ee
Therefore, in terms of the anisotropies \eqs(\ref{eq def delta EB}) and~(\ref{eq def delta ferm}), the pressure difference becomes
\be
p_x-p_z = p_y-p_z = - (\zeta_g+\hat{\zeta}_g) \cdot \left[A(\bna)-A(\ena)\right] - \zeta_f \cdot\sum_f A(\F_f),
\ee
as used in the body of the paper, \eq(\ref{eq:openaniso}). 
Note that at zero temperature, the perpendicular components coincide, and hence $A(\bna)-A(\ena)$ is proportional to $\trace\ena^2_{\parallel}(n)-\trace
\bna^2_{\parallel}(n)$.

We remark that the anisotropy coefficients and the multiplicative factors in \eq(\ref{eq:dlogzdxi}) are not completely independent. Consider the combination $I=\epsilon-p_x-p_y-p_z$, which can also be measured without introducing any anisotropy, as in \app\ref{sec imp}. Using \eqs(\ref{eq:edef})--(\ref{eq:dlogzdxi}) for the individual terms, we get for this combination at $\xi=1$,
\be
I = \frac{T}{V}  \bigg[
2(\zeta_g - \hat{\zeta}_g )\beta\cdot \expv{-S_g} + 
4\sum_f\frac{\partial (am_f)}{\partial \xi}\expv{\bar\Psi_f\Psi_f} -
\zeta_f\sum_fam_f\expv{\bar\Psi_f\Psi_f} \bigg],
\label{eq:eppp}
\ee
where we used that
\be
\sum_\mu \bar\Psi_f \slashed{D}_\mu \Psi_f = \Trace \frac{a\slashed{D}}{a\slashed{D}+am_f}
= \Trace \mathds{1} - am_f \,\bar\Psi_f\Psi_f,
\quad\quad\quad
\Trace \mathds{1} = N_s^3N_tN_c.
\label{eq:aux2}
\ee
Note that the constant term $N_s^3N_tN_c/4$ in \eq(\ref{eq:dlogzdxi}) is necessary to obtain a trace anomaly that does not contain any trivial constant shifts,
so that $I$ vanishes in the massless non-interacting case, as it should. The isotropic definition of the trace anomaly, \eq(\ref{eq_pre}), must equal \eq(\ref{eq:eppp}), which leads to the consistency relations
\be
2(\zeta_g-\hat{\zeta}_g)\beta = -\frac{\partial \beta}{\partial \log a}, \quad\quad\quad
4\frac{\partial \log(am_f)}{\partial \xi}-\zeta_f = -\frac{\partial\log(am_f)}{\partial \log a}.
\ee
The sum of the gluonic anisotropy coefficients is therefore related to the lattice $\beta$-function~\cite{Karsch:1982ve}. The two contributions were determined in perturbation theory for the pure gauge case, both for the Wilson gauge action~\cite{Karsch:1982ve} and for improved actions~\cite{Sakai:2000jm}. In the fermionic sector, the lattice $\gamma$-function shows up, and connects the $\xi$-dependence of $\xi_{f0}$ and that of $am_f$. 

\section{Euler-Heisenberg effective action}
\label{sec EH}

To enable a comparison of our non-perturbative results on the field strength anisotropy to perturbative expectations, in this appendix we compute the effective action of quarks coupled to a constant magnetic background field, plus constant chromo-electric and chromo-magnetic background fields to bi-quadratic order in the fields. This is a generalization of the celebrated Euler-Heisenberg action, in Euclidean space (and at zero temperature).

The dynamics of gauge fields in the path integral is governed by the sum of the gluonic action \eq(\ref{eq_gluonic_action_cont}) and the effective fermionic action, 
\begin{equation}
 \seff=-\log \det(\slashed{D}+m),
\end{equation}
which is obtained after integrating out fermions. For simplicity, we consider here a single fermion flavor with mass $m$ and charge $q$. 
For constant field strengths in QED, $\seff$ has been computed by Euler, Heisenberg and Schwinger~\cite{Heisenberg:1935qt,Schwinger:1951nm} to all orders. For a constant magnetic field $\ba$ in Euclidean space, it reads\footnote{In our conventions, the effective action corresponds to the free energy of the system, i.e. $\seff=-\log\Z$. Moreover, for the case of quarks, an additional factor of $N_c=3$ is to be included in $\seff$.},
\begin{equation}
 \seff(\ba)=\frac{V_4}{8\pi^2}\,q\ba\, m^2\int_0^\infty\frac{\d s}{s^2}\,e^{-s }\coth(q\ba s/m^2),
\label{eq all b}
\end{equation}
where $V_4=T/V$ again denotes the four-volume,
$q$ is the charge of the fermion (the electron), for a review see \cite{Dunne:2004nc}. This action is divergent for small $s$, i.e.\ in the UV, since 
$\coth(q\ba s/m^2)=m^2/q\ba s+q\ba s/(3m^2)-(q\ba s)^3/(45m^6)$.
The leading singularity is independent of $\ba$, and is thus absent in the difference $\seff(\ba)-\seff(0)$. The singularity quadratic in $\ba$ is taken care of by charge renormalization. The first non-trivial order is quartic, where, in diagrammatic language, four external photon legs interact via an electron loop (light-by-light scattering). The next order comes with an additional factor of $(q\ba)^2/m^4$, and is thus negligible for weak fields.

The fourth order term for constant field strength $G_{\mu\nu}$ in an arbitrary gauge group has been given by Novikov et al. \cite{Novikov:1983gd},
\begin{equation}
 \seff^{(4)}(G_{\mu\nu})=-\frac{V_4}{576\pi^2}\frac{\lambda^4}{m^4}\left[
  (G_{\mu\nu}G_{\mu\nu})^2-\frac{7}{10}\{G_{\mu\alpha},G_{\alpha\nu}\}^2-\frac{29}{70}
 [G_{\mu\alpha},G_{\alpha\nu}]^2+\frac{8}{35}[G_{\mu\nu},G_{\alpha\beta}]^2\right],
\label{eq fourth general}
\end{equation}
where $\lambda$ is the coupling in the covariant derivative $D_{\mu}=\partial_\mu+i\lambda A^G_\mu$ and $\{..,..\}$ and $[..,..]$ denote the anti-commutator and the commutator, respectively. For pure QED, upon replacing $\lambda G_{\mu\nu}\to q\fa_{\mu\nu}$, this yields 
\begin{equation}
 \seff^{(4)}(\ea,\ba)=-\frac{V_4}{360\pi^2}\frac{q^4}{m^4}\left[(\ve{\ea}^2+\ve{\ba}^2)^2-7(\ve{\ea}\,\ve{\ba})^2\right],
\label{eq fourth EH}
\end{equation}
reproducing the result by Euler and Heisenberg, $(-\ve{\ea}^2+\ve{\ba}^2)^2+7(\ve{\ea}\,\ve{\ba})^2$, if we change from Minkowski to Euclidean space, by multiplying the electric field by an imaginary unit. 

In the following, we again denote $\mathrm{SU}(3)$ fields by calligraphic letters and $\mathrm{U}(1)$ fields by straight characters. 
For QCD in external magnetic fields, one has to replace $\lambda G_{\mu\nu} \to \fna_{\mu\nu}+q\ba (\delta_{\mu 1}\delta_{\nu 2}-\delta_{\nu 1}\delta_{\mu 2})$ and a careful evaluation of \eq(\ref{eq fourth general}) to bi-quadratic order\footnote{We note that additional terms of the form $q\ba\,\trace(\bna\pa \fna_{\mu\nu}^2)$ and $q\ba\,\trace(\ena\pa \fna_{\mu\nu}\tilde{\fna}_{\mu\nu})$ also appear, and contribute to $\seff^{(3,1)}$. For instance, \eq(\ref{eq fourth Bpa}) below contains a term $\sum_a q\ba \,\bna_a^3=q\ba\, \trace \bna\pa^3$.} yields
 \begin{equation}
 \seff^{(2,2)}(\fna_{\mu\nu};\ba)=-\frac{V_4}{180\pi^2}\frac{(q\ba)^2}{m^4}\left[3\trace \bna\pa^2+\trace \bna\pe^2+\trace \ena\pe^2-\frac{5}{2}\trace \ena\pa^2\right],
\label{eq fourth final}
\end{equation}
in terms of the field strength components defined in \secref{sec observables gluonic aniso}. No topological charge term $\boldsymbol{\ena}\boldsymbol{\bna}$ appears, as expected from CP arguments in a purely magnetic external field. 

Thus, in perturbation theory for constant fields $|q\ba|, |\fna_{\mu\nu}|\ll m^2$ the chromo-electric field parallel to the external field has an increased action compared to the perpendicular fields, whereas the parallel chromo-magnetic field reduces the action. This means that parallel $E$-fields are disfavored, while parallel $B$-fields are favored. This is in qualitative agreement with our non-perturbative findings that $\dena>0$ and $\dbna<0$. 

The remainder of this appendix is devoted to check the main formula \eq(\ref{eq fourth final}). First, let us revisit the Abelian theory, by removing the traces and replacing 
calligraphic letters by $q$ times straight ones. This should be the fourth order result of Euler and Heisenberg, \eq(\ref{eq fourth EH}), up to the fact that here we have split the $B$-field in the $z$-direction artificially into $\ba\pa+\ba$ and computed only the terms of $\mathcal{O}(\ba^2)$. If we do the same in \eq(\ref{eq fourth EH}), we obtain in this order 
\begin{align}
 &-\frac{V_4}{360\pi^2}\frac{q^4}{m^4}\left[(\ea\pe^2+\ea\pa^2+\ba\pe^2+[\ba\pa+\ba]^2)^2-7(\ea\pe \ba\pe+\ea\pa[\ba\pa+\ba])^2\right]\\
 =&-\frac{V_4}{180\pi^2}\frac{q^4}{m^4}\,\ba^2\,
\frac{
 2\ea\pe^2+2\ea\pa^2+2\ba\pe^2+2\ba\pa^2+4\ba\pa^2-7\ea\pa^2}
 {2}+\ldots
\end{align}
which is what we get from \eq(\ref{eq fourth final}), too.

For the second check of \eq(\ref{eq fourth final}), we note that to this order it suffices to consider the effective action of just a constant field $\bna\pa$, or just a constant perpendicular field, or just a constant field $\ena\pa$, respectively, plus the external field $\ba$. In such a situation, the color
fields can be gauge transformed to a diagonal form, e.g.\ $\bna\pa=\mbox{diag}(\bna_1,\bna_2,\bna_3)$. 
Now the effective action splits into the sum over color sectors, in which one can use results for Abelian fields. Let us discuss the three cases separately.

For the case of $\bna\pa$, one has to replace $q\ba\to \bna_a+q\ba$ in \eq(\ref{eq all b}), with $a=1,2,3$. To fourth order, this yields
\begin{align}
 \seff^{(4)}(\bna_a;\ba)=\frac{V_4}{8\pi^2}\sum_a(\bna_a+q\ba)m^2\int_0^\infty\frac{ds}{s^2}\,e^{-s}\left(-\frac{s^3}{45m^6}\right)\cdot(\bna_a+q\ba)^3,
\label{eq fourth Bpa}
\end{align}
and to bi-quadratic order, using $\sum_a \bna_a^2=\trace \bna\pa^2$,
\begin{align}
  \seff^{(2,2)}(\bna\pa;\ba)=\frac{V_4}{8\pi^2}\sum_a\binom{4}{2}(q\ba)^2 \bna_a^2\frac{1}{m^4}\left(-\frac{1}{45}\right)
  =-\frac{V_4}{60\pi^2}\frac{(q\ba)^2}{m^4}\,\trace \bna\pa^2,
\end{align}
which agrees with the first term in \eq(\ref{eq fourth final}).

This result can be re-used for the perpendicular case. Electric fields can be turned into magnetic fields and vice versa, through the appropriate Lorentz-transformations. Here, we have a magnetic field (including charge) of $q\ba$ in the $z$-direction and perpendicular electric fields $\ena_a$ in, say, the $x$-direction. A `Euclidean boost' can be used to remove the electric field, leaving a magnetic field $\sqrt{\ena_a^2+(q\ba)^2}$ in the $z$-direction. The magnitude is fixed by preserving the Lorentz invariant combination $\ve{\ea}^2+\ve{\ba}^2$ (and $\ve{\ea}\ve{\ba}=0$). Thus, in the previous formula~\eq(\ref{eq fourth Bpa}) we have to replace $\bna_a+q\ba\to \sqrt{\ena_a^2+(q\ba)^2}$, arriving at
\begin{align}
 \seff^{(2,2)}(\ena\pe;\ba)=\frac{V_4}{8\pi^2}\sum_a\binom{2}{1}(q\ba)^2 \ena_a^2\frac{1}{m^4}\left(-\frac{1}{45}\right)
  =-\frac{V_4}{180\pi^2}\frac{(q\ba)^2}{m^4}\,\trace \ena\pe^2,
\end{align}
which applies to perpendicular magnetic fields, too.

For the $\ena\pa$ case, one needs to consider a magnetic field $q\ba$ and electric fields $\ena_a$ (again including charge factors) in the $z$-direction. The corresponding effective action can be obtained in analogy to \eq(\ref{eq all b}),
\begin{equation}
 S_a=\frac{V_4}{8\pi^2}\,\ena_aq\ba\int_0^\infty\frac{ds}{s}\,e^{-s}\coth(s\ena_a/m^2)\coth(sq\ba/m^2).
\label{eq sa e}
\end{equation}
Upon sending $\ena_a\to 0$, this reduces to \eq(\ref{eq all b}), as it should. For the bi-quadratic order it yields
\begin{align}
\seff^{(2,2)}(\ena\pa;\ba)
 &=\frac{V_4}{8\pi^2}\sum_a \ena_aq\ba\int_0^\infty\frac{ds}{s}\,e^{-s}\frac{s}{3m^2}\,\ena_a\frac{s}{3m^2}\,q\ba\nonumber\\
  &=\frac{V_4}{8\pi^2}\sum_a(q\ba)^2 \ena_a^2\frac{1}{m^4}\left(+\frac{1}{9}\right)
  =\frac{V_4}{72\pi^2}\frac{(q\ba)^2}{m^4}\,\trace \ena\pa^2,
\end{align}
which confirms the remaining term of \eq(\ref{eq fourth final}).

\bibliographystyle{JHEP_mcite}
\bibliography{aniso}

\ifx\mcitethebibliography\mciteundefinedmacro
\PackageError{unsrtM.bst}{mciteplus.sty has not been loaded}
{This bibstyle requires the use of the mciteplus package.}\fi
\begin{mcitethebibliography}{10}

\bibitem{Vachaspati:1991nm}
T.~Vachaspati, {\it {Magnetic fields from cosmological phase transitions}},
  {\em Phys. Lett. B} {\bf 265} (1991) 258\relax
\mciteBstWouldAddEndPuncttrue
\mciteSetBstMidEndSepPunct{\mcitedefaultmidpunct}
{\mcitedefaultendpunct}{\mcitedefaultseppunct}\relax
\EndOfBibitem
\bibitem{Duncan:1992hi}
R.~C. Duncan and C.~Thompson, {\it {Formation of very strongly magnetized
  neutron stars - implications for gamma-ray bursts}},  {\em Astrophys. J.}
  {\bf 392} (1992) L9\relax
\mciteBstWouldAddEndPuncttrue
\mciteSetBstMidEndSepPunct{\mcitedefaultmidpunct}
{\mcitedefaultendpunct}{\mcitedefaultseppunct}\relax
\EndOfBibitem
\bibitem{Skokov:2009qp}
V.~Skokov, A.~Y. Illarionov, and V.~Toneev, {\it {Estimate of the magnetic
  field strength in heavy-ion collisions}},  {\em Int. J. Mod. Phys. A} {\bf
  24} (2009) 5925, [\href{http://xxx.lanl.gov/abs/0907.1396}{{\tt
  arXiv:0907.1396}}]\relax
\mciteBstWouldAddEndPuncttrue
\mciteSetBstMidEndSepPunct{\mcitedefaultmidpunct}
{\mcitedefaultendpunct}{\mcitedefaultseppunct}\relax
\EndOfBibitem
\bibitem{Kharzeev:2007jp}
D.~E. Kharzeev, L.~D. McLerran, and H.~J. Warringa, {\it {The Effects of
  topological charge change in heavy ion collisions: `Event by event P and CP
  violation'}},  {\em Nucl.~Phys.} {\bf A803} (2008) 227,
  [\href{http://xxx.lanl.gov/abs/0711.0950}{{\tt arXiv:0711.0950}}]\relax
\mciteBstWouldAddEndPuncttrue
\mciteSetBstMidEndSepPunct{\mcitedefaultmidpunct}
{\mcitedefaultendpunct}{\mcitedefaultseppunct}\relax
\EndOfBibitem
\bibitem{Fukushima:2008xe}
K.~Fukushima, D.~E. Kharzeev, and H.~J. Warringa, {\it {The Chiral Magnetic
  Effect}},  {\em Phys.~Rev.} {\bf D78} (2008) 074033,
  [\href{http://xxx.lanl.gov/abs/0808.3382}{{\tt arXiv:0808.3382}}]\relax
\mciteBstWouldAddEndPuncttrue
\mciteSetBstMidEndSepPunct{\mcitedefaultmidpunct}
{\mcitedefaultendpunct}{\mcitedefaultseppunct}\relax
\EndOfBibitem
\bibitem{Gusynin:1995nb}
V.~P. Gusynin, V.~A. Miransky, and I.~A. Shovkovy, {\it {Dimensional reduction
  and catalysis of dynamical symmetry breaking by a magnetic field}},  {\em
  Nucl. Phys. B} {\bf 462} (1996) 249,
  [\href{http://xxx.lanl.gov/abs/hep-ph/9509320}{{\tt hep-ph/9509320}}]\relax
\mciteBstWouldAddEndPuncttrue
\mciteSetBstMidEndSepPunct{\mcitedefaultmidpunct}
{\mcitedefaultendpunct}{\mcitedefaultseppunct}\relax
\EndOfBibitem
\bibitem{Landau:1930}
L.~Landau, {\it {Diamagnetismus der Metalle}},  {\em Zeitschrift f{\"u}r Physik
  A, Hadrons and Nuclei} {\bf 64} (1930) 629\relax
\mciteBstWouldAddEndPuncttrue
\mciteSetBstMidEndSepPunct{\mcitedefaultmidpunct}
{\mcitedefaultendpunct}{\mcitedefaultseppunct}\relax
\EndOfBibitem
\bibitem{Banks:1979yr}
T.~Banks and A.~Casher, {\it {Chiral Symmetry Breaking in Confining Theories}},
   {\em Nucl.~Phys.} {\bf B169} (1980) 103\relax
\mciteBstWouldAddEndPuncttrue
\mciteSetBstMidEndSepPunct{\mcitedefaultmidpunct}
{\mcitedefaultendpunct}{\mcitedefaultseppunct}\relax
\EndOfBibitem
\bibitem{Endrodi:2013cs}
G.~Endr\H{o}di, {\it {QCD equation of state at nonzero magnetic fields in the
  Hadron Resonance Gas model}},  {\em JHEP} {\bf 1304} (2013) 023,
  [\href{http://xxx.lanl.gov/abs/1301.1307}{{\tt arXiv:1301.1307}}]\relax
\mciteBstWouldAddEndPuncttrue
\mciteSetBstMidEndSepPunct{\mcitedefaultmidpunct}
{\mcitedefaultendpunct}{\mcitedefaultseppunct}\relax
\EndOfBibitem
\bibitem{D'Elia:2010nq}
M.~D'Elia, S.~Mukherjee, and F.~Sanfilippo, {\it {QCD Phase Transition in a
  Strong Magnetic Background}},  {\em Phys. Rev. D} {\bf 82} (2010) 051501,
  [\href{http://xxx.lanl.gov/abs/1005.5365}{{\tt arXiv:1005.5365}}]\relax
\mciteBstWouldAddEndPuncttrue
\mciteSetBstMidEndSepPunct{\mcitedefaultmidpunct}
{\mcitedefaultendpunct}{\mcitedefaultseppunct}\relax
\EndOfBibitem
\bibitem{Bruckmann:2011zx}
F.~Bruckmann and G.~Endr\H{o}di, {\it {Dressed Wilson loops as dual condensates
  in response to magnetic and electric fields}},  {\em Phys.~Rev.} {\bf D84}
  (2011) 074506, [\href{http://xxx.lanl.gov/abs/1104.5664}{{\tt
  arXiv:1104.5664}}]\relax
\mciteBstWouldAddEndPuncttrue
\mciteSetBstMidEndSepPunct{\mcitedefaultmidpunct}
{\mcitedefaultendpunct}{\mcitedefaultseppunct}\relax
\EndOfBibitem
\bibitem{Bali:2011qj}
G.~Bali, F.~Bruckmann, G.~Endr\H{o}di, Z.~Fodor, S.~Katz, et~al., {\it {The QCD
  phase diagram for external magnetic fields}},  {\em JHEP} {\bf 1202} (2012)
  044, [\href{http://xxx.lanl.gov/abs/1111.4956}{{\tt arXiv:1111.4956}}]\relax
\mciteBstWouldAddEndPuncttrue
\mciteSetBstMidEndSepPunct{\mcitedefaultmidpunct}
{\mcitedefaultendpunct}{\mcitedefaultseppunct}\relax
\EndOfBibitem
\bibitem{Bali:2012zg}
G.~Bali, F.~Bruckmann, G.~Endr\H{o}di, Z.~Fodor, S.~Katz, et~al., {\it {QCD
  quark condensate in external magnetic fields}},  {\em Phys.~Rev.} {\bf D86}
  (2012) 071502, [\href{http://xxx.lanl.gov/abs/1206.4205}{{\tt
  arXiv:1206.4205}}]\relax
\mciteBstWouldAddEndPuncttrue
\mciteSetBstMidEndSepPunct{\mcitedefaultmidpunct}
{\mcitedefaultendpunct}{\mcitedefaultseppunct}\relax
\EndOfBibitem
\bibitem{Ilgenfritz:2012fw}
E.-M. Ilgenfritz, M.~Kalinowski, M.~M{\"u}ller-Preussker, B.~Petersson, and
  A.~Schreiber, {\it {Two-color QCD with staggered fermions at finite
  temperature under the influence of a magnetic field}},  {\em Phys. Rev.} {\bf
  D85} (2012) 114504, [\href{http://xxx.lanl.gov/abs/1203.3360}{{\tt
  arXiv:1203.3360}}]\relax
\mciteBstWouldAddEndPuncttrue
\mciteSetBstMidEndSepPunct{\mcitedefaultmidpunct}
{\mcitedefaultendpunct}{\mcitedefaultseppunct}\relax
\EndOfBibitem
\bibitem{Shovkovy:2012zn}
I.~A. Shovkovy, {\it {Magnetic Catalysis: A Review}},
  \href{http://xxx.lanl.gov/abs/1207.5081}{{\tt arXiv:1207.5081}}\relax
\mciteBstWouldAddEndPuncttrue
\mciteSetBstMidEndSepPunct{\mcitedefaultmidpunct}
{\mcitedefaultendpunct}{\mcitedefaultseppunct}\relax
\EndOfBibitem
\bibitem{Bruckmann:2013oba}
F.~Bruckmann, G.~Endr\H{o}di, and T.~Kov\'acs, {\it {Inverse magnetic catalysis
  and the Polyakov loop}},  {\em JHEP} {\bf 1304} (2013) 112,
  [\href{http://xxx.lanl.gov/abs/1303.3972}{{\tt arXiv:1303.3972}}]\relax
\mciteBstWouldAddEndPuncttrue
\mciteSetBstMidEndSepPunct{\mcitedefaultmidpunct}
{\mcitedefaultendpunct}{\mcitedefaultseppunct}\relax
\EndOfBibitem
\bibitem{Basar:2011by}
G.~Ba{\c s}ar, G.~V. Dunne, and D.~E. Kharzeev, {\it {Electric dipole moment
  induced by a QCD instanton in an external magnetic field}},  {\em Phys.~Rev.}
  {\bf D85} (2012) 045026, [\href{http://xxx.lanl.gov/abs/1112.0532}{{\tt
  arXiv:1112.0532}}]\relax
\mciteBstWouldAddEndPuncttrue
\mciteSetBstMidEndSepPunct{\mcitedefaultmidpunct}
{\mcitedefaultendpunct}{\mcitedefaultseppunct}\relax
\EndOfBibitem
\bibitem{Ioffe:1983ju}
B.~Ioffe and A.~V. Smilga, {\it {Nucleon Magnetic Moments and Magnetic
  Properties of Vacuum in QCD}},  {\em Nucl.~Phys.} {\bf B232} (1984) 109\relax
\mciteBstWouldAddEndPuncttrue
\mciteSetBstMidEndSepPunct{\mcitedefaultmidpunct}
{\mcitedefaultendpunct}{\mcitedefaultseppunct}\relax
\EndOfBibitem
\bibitem{Bali:2012jv}
G.~Bali, F.~Bruckmann, M.~Constantinou, M.~Costa, G.~Endr\H{o}di, et~al., {\it
  {Magnetic susceptibility of QCD at zero and at finite temperature from the
  lattice}},  {\em Phys.~Rev.} {\bf D86} (2012) 094512,
  [\href{http://xxx.lanl.gov/abs/1209.6015}{{\tt arXiv:1209.6015}}]\relax
\mciteBstWouldAddEndPuncttrue
\mciteSetBstMidEndSepPunct{\mcitedefaultmidpunct}
{\mcitedefaultendpunct}{\mcitedefaultseppunct}\relax
\EndOfBibitem
\bibitem{Buividovich:2009ih}
P.~Buividovich, M.~Chernodub, E.~Luschevskaya, and M.~Polikarpov, {\it {Chiral
  magnetization of non-Abelian vacuum: A Lattice study}},  {\em Nucl.~Phys.}
  {\bf B826} (2010) 313, [\href{http://xxx.lanl.gov/abs/0906.0488}{{\tt
  arXiv:0906.0488}}]\relax
\mciteBstWouldAddEndPuncttrue
\mciteSetBstMidEndSepPunct{\mcitedefaultmidpunct}
{\mcitedefaultendpunct}{\mcitedefaultseppunct}\relax
\EndOfBibitem
\bibitem{Braguta:2010ej}
V.~Braguta, P.~Buividovich, T.~Kalaydzhyan, S.~Kuznetsov, and M.~Polikarpov,
  {\it {The Chiral Magnetic Effect and chiral symmetry breaking in SU(3)
  quenched lattice gauge theory}},  {\em Phys.~Atom.~Nucl.} {\bf 75} (2012)
  488, [\href{http://xxx.lanl.gov/abs/1011.3795}{{\tt arXiv:1011.3795}}]\relax
\mciteBstWouldAddEndPuncttrue
\mciteSetBstMidEndSepPunct{\mcitedefaultmidpunct}
{\mcitedefaultendpunct}{\mcitedefaultseppunct}\relax
\EndOfBibitem
\bibitem{Blandford:1982}
R.~Blandford and L.~Hernquist, {\it {Magnetic susceptibility of a neutron star
  crust}},  {\em J. Phys} {\bf C15} (1982) 6233\relax
\mciteBstWouldAddEndPuncttrue
\mciteSetBstMidEndSepPunct{\mcitedefaultmidpunct}
{\mcitedefaultendpunct}{\mcitedefaultseppunct}\relax
\EndOfBibitem
\bibitem{Ferrer:2010wz}
E.~J. Ferrer, V.~de~la Incera, J.~P. Keith, I.~Portillo, and P.~P. Springsteen,
  {\it {Equation of State of a Dense and Magnetized Fermion System}},  {\em
  Phys.~Rev.} {\bf C82} (2010) 065802,
  [\href{http://xxx.lanl.gov/abs/1009.3521}{{\tt arXiv:1009.3521}}]\relax
\mciteBstWouldAddEndPuncttrue
\mciteSetBstMidEndSepPunct{\mcitedefaultmidpunct}
{\mcitedefaultendpunct}{\mcitedefaultseppunct}\relax
\EndOfBibitem
\bibitem{Potekhin:2011eb}
A.~Potekhin and D.~Yakovlev, {\it {Comment on `Equation of state of dense and
  magnetized fermion system'}},  {\em Phys.~Rev.} {\bf C85} (2012) 039801,
  [\href{http://xxx.lanl.gov/abs/1109.3783}{{\tt arXiv:1109.3783}}]\relax
\mciteBstWouldAddEndPuncttrue
\mciteSetBstMidEndSepPunct{\mcitedefaultmidpunct}
{\mcitedefaultendpunct}{\mcitedefaultseppunct}\relax
\EndOfBibitem
\bibitem{Ferrer:2011ze}
E.~J. Ferrer and V.~de~la Incera, {\it {Reply to comment on `Equation of state
  of dense and magnetized fermion system'}},  {\em Phys.~Rev.} {\bf C85} (2012)
  039802, [\href{http://xxx.lanl.gov/abs/1110.0420}{{\tt
  arXiv:1110.0420}}]\relax
\mciteBstWouldAddEndPuncttrue
\mciteSetBstMidEndSepPunct{\mcitedefaultmidpunct}
{\mcitedefaultendpunct}{\mcitedefaultseppunct}\relax
\EndOfBibitem
\bibitem{Strickland:2012vu}
M.~Strickland, V.~Dexheimer, and D.~Menezes, {\it {Bulk Properties of a Fermi
  Gas in a Magnetic Field}},  {\em Phys. Rev.} {\bf D86} (2012) 125032,
  [\href{http://xxx.lanl.gov/abs/1209.3276}{{\tt arXiv:1209.3276}}]\relax
\mciteBstWouldAddEndPuncttrue
\mciteSetBstMidEndSepPunct{\mcitedefaultmidpunct}
{\mcitedefaultendpunct}{\mcitedefaultseppunct}\relax
\EndOfBibitem
\bibitem{Kandus:2010nw}
A.~Kandus, K.~E. Kunze, and C.~G. Tsagas, {\it {Primordial magnetogenesis}},
  {\em Phys.Rept.} {\bf 505} (2011) 1--58,
  [\href{http://xxx.lanl.gov/abs/1007.3891}{{\tt arXiv:1007.3891}}]\relax
\mciteBstWouldAddEndPuncttrue
\mciteSetBstMidEndSepPunct{\mcitedefaultmidpunct}
{\mcitedefaultendpunct}{\mcitedefaultseppunct}\relax
\EndOfBibitem
\bibitem{Muller:2012zq}
B.~M{\"u}ller, J.~Schukraft, and B.~Wyslouch, {\it {First Results from Pb+Pb
  collisions at the LHC}},  {\em Ann.Rev.Nucl.Part.Sci.} {\bf 62} (2012)
  361--386, [\href{http://xxx.lanl.gov/abs/1202.3233}{{\tt
  arXiv:1202.3233}}]\relax
\mciteBstWouldAddEndPuncttrue
\mciteSetBstMidEndSepPunct{\mcitedefaultmidpunct}
{\mcitedefaultendpunct}{\mcitedefaultseppunct}\relax
\EndOfBibitem
\bibitem{Schenke:2012wb}
B.~Schenke, P.~Tribedy, and R.~Venugopalan, {\it {Fluctuating Glasma initial
  conditions and flow in heavy ion collisions}},  {\em Phys.Rev.Lett.} {\bf
  108} (2012) 252301, [\href{http://xxx.lanl.gov/abs/1202.6646}{{\tt
  arXiv:1202.6646}}]\relax
\mciteBstWouldAddEndPuncttrue
\mciteSetBstMidEndSepPunct{\mcitedefaultmidpunct}
{\mcitedefaultendpunct}{\mcitedefaultseppunct}\relax
\EndOfBibitem
\bibitem{Heinz:2013th}
U.~W. Heinz and R.~Snellings, {\it {Collective flow and viscosity in
  relativistic heavy-ion collisions}},
  \href{http://xxx.lanl.gov/abs/1301.2826}{{\tt arXiv:1301.2826}}\relax
\mciteBstWouldAddEndPuncttrue
\mciteSetBstMidEndSepPunct{\mcitedefaultmidpunct}
{\mcitedefaultendpunct}{\mcitedefaultseppunct}\relax
\EndOfBibitem
\bibitem{Borsanyi:2010cj}
S.~Bors\'anyi, G.~Endr\H{o}di, Z.~Fodor, A.~Jakov\'ac, S.~D. Katz, et~al., {\it
  {The QCD equation of state with dynamical quarks}},  {\em JHEP} {\bf 1011}
  (2010) 077, [\href{http://xxx.lanl.gov/abs/1007.2580}{{\tt
  arXiv:1007.2580}}]\relax
\mciteBstWouldAddEndPuncttrue
\mciteSetBstMidEndSepPunct{\mcitedefaultmidpunct}
{\mcitedefaultendpunct}{\mcitedefaultseppunct}\relax
\EndOfBibitem
\bibitem{Weisz:1982zw}
P.~Weisz, {\it {Continuum Limit Improved Lattice Action for Pure Yang-Mills
  Theory. 1.}},  {\em Nucl.~Phys.} {\bf B212} (1983) 1\relax
\mciteBstWouldAddEndPuncttrue
\mciteSetBstMidEndSepPunct{\mcitedefaultmidpunct}
{\mcitedefaultendpunct}{\mcitedefaultseppunct}\relax
\EndOfBibitem
\bibitem{Moran:2008ra}
P.~J. Moran and D.~B. Leinweber, {\it {Over-improved stout-link smearing}},
  {\em Phys.~Rev.} {\bf D77} (2008) 094501,
  [\href{http://xxx.lanl.gov/abs/0801.1165}{{\tt arXiv:0801.1165}}]\relax
\mciteBstWouldAddEndPuncttrue
\mciteSetBstMidEndSepPunct{\mcitedefaultmidpunct}
{\mcitedefaultendpunct}{\mcitedefaultseppunct}\relax
\EndOfBibitem
\bibitem{BilsonThompson:2002jk}
S.~O. Bilson-Thompson, D.~B. Leinweber, and A.~G. Williams, {\it {Highly
  improved lattice field strength tensor}},  {\em Annals Phys.} {\bf 304}
  (2003) 1, [\href{http://xxx.lanl.gov/abs/hep-lat/0203008}{{\tt
  hep-lat/0203008}}]\relax
\mciteBstWouldAddEndPuncttrue
\mciteSetBstMidEndSepPunct{\mcitedefaultmidpunct}
{\mcitedefaultendpunct}{\mcitedefaultseppunct}\relax
\EndOfBibitem
\bibitem{Ilgenfritz:2008ia}
E.-M. Ilgenfritz, D.~Leinweber, P.~Moran, K.~Koller, G.~Schierholz, et~al.,
  {\it {Vacuum structure revealed by over-improved stout-link smearing compared
  with the overlap analysis for quenched QCD}},  {\em Phys. Rev.} {\bf D77}
  (2008) 074502, [\href{http://xxx.lanl.gov/abs/0801.1725}{{\tt
  arXiv:0801.1725}}]\relax
\mciteBstWouldAddEndPuncttrue
\mciteSetBstMidEndSepPunct{\mcitedefaultmidpunct}
{\mcitedefaultendpunct}{\mcitedefaultseppunct}\relax
\EndOfBibitem
\bibitem{Bruckmann:2011ve}
F.~Bruckmann, F.~Gruber, N.~Cundy, A.~Sch{\"a}fer, and T.~Lippert, {\it
  {Topology of dynamical lattice configurations including results from
  dynamical overlap fermions}},  {\em Phys.~Lett.} {\bf B707} (2012) 278,
  [\href{http://xxx.lanl.gov/abs/1107.0897}{{\tt arXiv:1107.0897}}]\relax
\mciteBstWouldAddEndPuncttrue
\mciteSetBstMidEndSepPunct{\mcitedefaultmidpunct}
{\mcitedefaultendpunct}{\mcitedefaultseppunct}\relax
\EndOfBibitem
\bibitem{Niedermayer:1998bi}
F.~Niedermayer, {\it {Exact chiral symmetry, topological charge and related
  topics}},  {\em Nucl.~Phys.~Proc.~Suppl.} {\bf 73} (1999) 105,
  [\href{http://xxx.lanl.gov/abs/hep-lat/9810026}{{\tt hep-lat/9810026}}]\relax
\mciteBstWouldAddEndPuncttrue
\mciteSetBstMidEndSepPunct{\mcitedefaultmidpunct}
{\mcitedefaultendpunct}{\mcitedefaultseppunct}\relax
\EndOfBibitem
\bibitem{Horvath:2002yn}
I.~Horvath, S.~Dong, T.~Draper, F.~Lee, K.-F. Liu, et~al., {\it {On the local
  structure of topological charge fluctuations in QCD}},  {\em Phys.~Rev.} {\bf
  D67} (2003) 011501, [\href{http://xxx.lanl.gov/abs/hep-lat/0203027}{{\tt
  hep-lat/0203027}}]\relax
\mciteBstWouldAddEndPuncttrue
\mciteSetBstMidEndSepPunct{\mcitedefaultmidpunct}
{\mcitedefaultendpunct}{\mcitedefaultseppunct}\relax
\EndOfBibitem
\bibitem{Rafelski:1998tc}
J.~Rafelski, {\it {Electromagnetic fields in the QCD vacuum}},
  \href{http://xxx.lanl.gov/abs/hep-ph/9806389}{{\tt hep-ph/9806389}}\relax
\mciteBstWouldAddEndPuncttrue
\mciteSetBstMidEndSepPunct{\mcitedefaultmidpunct}
{\mcitedefaultendpunct}{\mcitedefaultseppunct}\relax
\EndOfBibitem
\bibitem{Elze:1998wm}
H.~T. Elze, B.~M{\"u}ller, and J.~Rafelski, {\it {Interfering QCD/QED vacuum
  polarization}},  \href{http://xxx.lanl.gov/abs/hep-ph/9811372}{{\tt
  hep-ph/9811372}}\relax
\mciteBstWouldAddEndPuncttrue
\mciteSetBstMidEndSepPunct{\mcitedefaultmidpunct}
{\mcitedefaultendpunct}{\mcitedefaultseppunct}\relax
\EndOfBibitem
\bibitem{D'Elia:2012zw}
M.~D'Elia, M.~Mariti, and F.~Negro, {\it {Susceptibility of the QCD vacuum to
  CP-odd electromagnetic background fields}},
  \href{http://xxx.lanl.gov/abs/1209.0722}{{\tt arXiv:1209.0722}}\relax
\mciteBstWouldAddEndPuncttrue
\mciteSetBstMidEndSepPunct{\mcitedefaultmidpunct}
{\mcitedefaultendpunct}{\mcitedefaultseppunct}\relax
\EndOfBibitem
\bibitem{Seiler:1987ig}
E.~Seiler and I.~Stamatescu, {\it {Some remarks on the Witten-Venziano formula
  for the eta-prime mass}},  \href{http://xxx.lanl.gov/abs/MPI-PAE/PTh
  10/87}{{\tt MPI-PAE/PTh 10/87}}\relax
\mciteBstWouldAddEndPuncttrue
\mciteSetBstMidEndSepPunct{\mcitedefaultmidpunct}
{\mcitedefaultendpunct}{\mcitedefaultseppunct}\relax
\EndOfBibitem
\bibitem{Ioffe:2002be}
B.~Ioffe and K.~Zyablyuk, {\it {Gluon condensate in charmonium sum rules with
  three loop corrections}},  {\em Eur.~Phys.~J.} {\bf C27} (2003) 229,
  [\href{http://xxx.lanl.gov/abs/hep-ph/0207183}{{\tt hep-ph/0207183}}]\relax
\mciteBstWouldAddEndPuncttrue
\mciteSetBstMidEndSepPunct{\mcitedefaultmidpunct}
{\mcitedefaultendpunct}{\mcitedefaultseppunct}\relax
\EndOfBibitem
\bibitem{landau1971classical}
L.~Landau and E.~Lifshits, {\em The classical theory of fields}.
\newblock Course on Theoretical Physics, Vol. 2. Pergamon Press, 1971\relax
\mciteBstWouldAddEndPuncttrue
\mciteSetBstMidEndSepPunct{\mcitedefaultmidpunct}
{\mcitedefaultendpunct}{\mcitedefaultseppunct}\relax
\EndOfBibitem
\bibitem{Canuto:1969ct}
V.~Canuto and H.~Chiu, {\it {Quantum theory of an electron gas in intense
  magnetic fields}},  {\em Phys. Rev.} {\bf 173} (1968) 1210\relax
\mciteBstWouldAddEndPuncttrue
\mciteSetBstMidEndSepPunct{\mcitedefaultmidpunct}
{\mcitedefaultendpunct}{\mcitedefaultseppunct}\relax
\EndOfBibitem
\bibitem{Martinez:2003dz}
A.~P. Martinez, H.~P. Rojas, and H.~J. Mosquera~Cuesta, {\it {Magnetic collapse
  of a neutron gas: Can magnetars indeed be formed?}},  {\em Eur.~Phys.~J.}
  {\bf C29} (2003) 111, [\href{http://xxx.lanl.gov/abs/astro-ph/0303213}{{\tt
  astro-ph/0303213}}]\relax
\mciteBstWouldAddEndPuncttrue
\mciteSetBstMidEndSepPunct{\mcitedefaultmidpunct}
{\mcitedefaultendpunct}{\mcitedefaultseppunct}\relax
\EndOfBibitem
\bibitem{jackson1975classical}
J.~Jackson, {\em Classical electrodynamics}.
\newblock Wiley, 1975\relax
\mciteBstWouldAddEndPuncttrue
\mciteSetBstMidEndSepPunct{\mcitedefaultmidpunct}
{\mcitedefaultendpunct}{\mcitedefaultseppunct}\relax
\EndOfBibitem
\bibitem{rohrlich2007classical}
F.~Rohrlich, {\em Classical Charged Particles}.
\newblock World Scientific, 2007\relax
\mciteBstWouldAddEndPuncttrue
\mciteSetBstMidEndSepPunct{\mcitedefaultmidpunct}
{\mcitedefaultendpunct}{\mcitedefaultseppunct}\relax
\EndOfBibitem
\bibitem{0022-3719-15-30-017}
R.~D. Blandford and L.~Hernquist, {\it Magnetic susceptibility of a neutron
  star crust},  {\em J. Phys. C: Solid State Physics} {\bf 15} (1982), no.~30
  6233\relax
\mciteBstWouldAddEndPuncttrue
\mciteSetBstMidEndSepPunct{\mcitedefaultmidpunct}
{\mcitedefaultendpunct}{\mcitedefaultseppunct}\relax
\EndOfBibitem
\bibitem{landau1995electrodynamics}
L.~Landau, E.~Lifshitz, and L.~Pitaevskii, {\em Electrodynamics of continuous
  media}.
\newblock Course of theoretical physics. Butterworth-Heinemann, 1995\relax
\mciteBstWouldAddEndPuncttrue
\mciteSetBstMidEndSepPunct{\mcitedefaultmidpunct}
{\mcitedefaultendpunct}{\mcitedefaultseppunct}\relax
\EndOfBibitem
\bibitem{Karsch:1982ve}
F.~Karsch, {\it {SU(N) Gauge Theory Couplings on Asymmetric Lattices}},  {\em
  Nucl. Phys.} {\bf B205} (1982) 285\relax
\mciteBstWouldAddEndPuncttrue
\mciteSetBstMidEndSepPunct{\mcitedefaultmidpunct}
{\mcitedefaultendpunct}{\mcitedefaultseppunct}\relax
\EndOfBibitem
\bibitem{Karsch:1989fu}
F.~Karsch and I.~Stamatescu, {\it {QCD thermodynamics with light quarks:
  Quantum corrections to the fermionic anisotropy parameter}},  {\em
  Phys.~Lett.} {\bf B227} (1989) 153\relax
\mciteBstWouldAddEndPuncttrue
\mciteSetBstMidEndSepPunct{\mcitedefaultmidpunct}
{\mcitedefaultendpunct}{\mcitedefaultseppunct}\relax
\EndOfBibitem
\bibitem{Fraga:2012ev}
E.~S. Fraga, J.~Noronha, and L.~F. Palhares, {\it {Large Nc Deconfinement
  Transition in the Presence of a Magnetic Field}},
  \href{http://xxx.lanl.gov/abs/1207.7094}{{\tt arXiv:1207.7094}}\relax
\mciteBstWouldAddEndPuncttrue
\mciteSetBstMidEndSepPunct{\mcitedefaultmidpunct}
{\mcitedefaultendpunct}{\mcitedefaultseppunct}\relax
\EndOfBibitem
\bibitem{Leutwyler:1992yt}
H.~Leutwyler and A.~V. Smilga, {\it {Spectrum of Dirac operator and role of
  winding number in QCD}},  {\em Phys.~Rev.} {\bf D46} (1992) 5607\relax
\mciteBstWouldAddEndPuncttrue
\mciteSetBstMidEndSepPunct{\mcitedefaultmidpunct}
{\mcitedefaultendpunct}{\mcitedefaultseppunct}\relax
\EndOfBibitem
\bibitem{Schwinger:1951nm}
J.~S. Schwinger, {\it {On gauge invariance and vacuum polarization}},  {\em
  Phys.~Rev.} {\bf 82} (1951) 664\relax
\mciteBstWouldAddEndPuncttrue
\mciteSetBstMidEndSepPunct{\mcitedefaultmidpunct}
{\mcitedefaultendpunct}{\mcitedefaultseppunct}\relax
\EndOfBibitem
\bibitem{Elmfors:1993bm}
P.~Elmfors, D.~Persson, and B.-S. Skagerstam, {\it {Real time thermal
  propagators and the QED effective action for an external magnetic field}},
  {\em Astropart. Phys.} {\bf 2} (1994) 299,
  [\href{http://xxx.lanl.gov/abs/hep-ph/9312226}{{\tt hep-ph/9312226}}]\relax
\mciteBstWouldAddEndPuncttrue
\mciteSetBstMidEndSepPunct{\mcitedefaultmidpunct}
{\mcitedefaultendpunct}{\mcitedefaultseppunct}\relax
\EndOfBibitem
\bibitem{Dunne:2004nc}
G.~V. Dunne, {\it {Heisenberg-Euler effective Lagrangians: Basics and
  extensions}},  \href{http://xxx.lanl.gov/abs/hep-th/0406216}{{\tt
  hep-th/0406216}}\relax
\mciteBstWouldAddEndPuncttrue
\mciteSetBstMidEndSepPunct{\mcitedefaultmidpunct}
{\mcitedefaultendpunct}{\mcitedefaultseppunct}\relax
\EndOfBibitem
\bibitem{Cheng:2007jq}
M.~Cheng, N.~Christ, S.~Datta, J.~van~der Heide, C.~Jung, et~al., {\it {The QCD
  equation of state with almost physical quark masses}},  {\em Phys.~Rev.} {\bf
  D77} (2008) 014511, [\href{http://xxx.lanl.gov/abs/0710.0354}{{\tt
  arXiv:0710.0354}}]\relax
\mciteBstWouldAddEndPuncttrue
\mciteSetBstMidEndSepPunct{\mcitedefaultmidpunct}
{\mcitedefaultendpunct}{\mcitedefaultseppunct}\relax
\EndOfBibitem
\bibitem{Lee:1993af}
W.-J. Lee, {\it {Quark mass renormalization on the lattice with staggered
  fermions}},  {\em Phys.~Rev.} {\bf D49} (1994) 3563,
  [\href{http://xxx.lanl.gov/abs/hep-lat/9310018}{{\tt hep-lat/9310018}}]\relax
\mciteBstWouldAddEndPuncttrue
\mciteSetBstMidEndSepPunct{\mcitedefaultmidpunct}
{\mcitedefaultendpunct}{\mcitedefaultseppunct}\relax
\EndOfBibitem
\bibitem{Engels:1981qx}
J.~Engels, F.~Karsch, H.~Satz, and I.~Montvay, {\it {Gauge Field Thermodynamics
  for the SU(2) Yang-Mills System}},  {\em Nucl. Phys.} {\bf B205} (1982)
  545\relax
\mciteBstWouldAddEndPuncttrue
\mciteSetBstMidEndSepPunct{\mcitedefaultmidpunct}
{\mcitedefaultendpunct}{\mcitedefaultseppunct}\relax
\EndOfBibitem
\bibitem{Bali:2000un}
G.~S. Bali, {\it {Casimir scaling of SU(3) static potentials}},  {\em
  Phys.~Rev.} {\bf D62} (2000) 114503,
  [\href{http://xxx.lanl.gov/abs/hep-lat/0006022}{{\tt hep-lat/0006022}}]\relax
\mciteBstWouldAddEndPuncttrue
\mciteSetBstMidEndSepPunct{\mcitedefaultmidpunct}
{\mcitedefaultendpunct}{\mcitedefaultseppunct}\relax
\EndOfBibitem
\bibitem{Borsanyi:2012zr}
S.~Bors\'anyi, S.~D{\"u}rr, Z.~Fodor, S.~D. Katz, S.~Krieg, et~al., {\it
  {Anisotropy tuning with the Wilson flow}},
  \href{http://xxx.lanl.gov/abs/1205.0781}{{\tt arXiv:1205.0781}}\relax
\mciteBstWouldAddEndPuncttrue
\mciteSetBstMidEndSepPunct{\mcitedefaultmidpunct}
{\mcitedefaultendpunct}{\mcitedefaultseppunct}\relax
\EndOfBibitem
\bibitem{Levkova:2006gn}
L.~Levkova, T.~Manke, and R.~Mawhinney, {\it {Two-flavor QCD thermodynamics
  using anisotropic lattices}},  {\em Phys.~Rev.} {\bf D73} (2006) 074504,
  [\href{http://xxx.lanl.gov/abs/hep-lat/0603031}{{\tt hep-lat/0603031}}]\relax
\mciteBstWouldAddEndPuncttrue
\mciteSetBstMidEndSepPunct{\mcitedefaultmidpunct}
{\mcitedefaultendpunct}{\mcitedefaultseppunct}\relax
\EndOfBibitem
\bibitem{Sakai:2000jm}
S.~Sakai, T.~Saito, and A.~Nakamura, {\it {Anisotropic lattice with improved
  gauge actions. 1. Study of fundamental parameters in weak coupling regions}},
   {\em Nucl.~Phys.} {\bf B584} (2000) 528,
  [\href{http://xxx.lanl.gov/abs/hep-lat/0002029}{{\tt hep-lat/0002029}}]\relax
\mciteBstWouldAddEndPuncttrue
\mciteSetBstMidEndSepPunct{\mcitedefaultmidpunct}
{\mcitedefaultendpunct}{\mcitedefaultseppunct}\relax
\EndOfBibitem
\bibitem{Heisenberg:1935qt}
W.~Heisenberg and H.~Euler, {\it {Consequences of Dirac's theory of
  positrons}},  {\em Z.~Phys.} {\bf 98} (1936) 714,
  [\href{http://xxx.lanl.gov/abs/physics/0605038}{{\tt physics/0605038}}]\relax
\mciteBstWouldAddEndPuncttrue
\mciteSetBstMidEndSepPunct{\mcitedefaultmidpunct}
{\mcitedefaultendpunct}{\mcitedefaultseppunct}\relax
\EndOfBibitem
\bibitem{Novikov:1983gd}
V.~Novikov, M.~A. Shifman, A.~Vainshtein, and V.~I. Zakharov, {\it
  {Calculations in External Fields in Quantum Chromodynamics. Technical
  Review}},  {\em Fortsch.~Phys.} {\bf 32} (1984) 585\relax
\mciteBstWouldAddEndPuncttrue
\mciteSetBstMidEndSepPunct{\mcitedefaultmidpunct}
{\mcitedefaultendpunct}{\mcitedefaultseppunct}\relax
\EndOfBibitem
\end{mcitethebibliography}

\end{document}